\documentclass[final,10pt,3p]{elsarticle}
\journal{Computer Physics Communication}
\usepackage{color}
\usepackage[dvipsnames,svgnames,table]{xcolor}

\usepackage{mathtools}
\usepackage{float}
\usepackage{booktabs}
\usepackage{graphicx}
\usepackage{dcolumn}
\usepackage{array}
\usepackage{algorithm}
\usepackage{algpseudocode}
\usepackage{lipsum}
\usepackage{bm}
\usepackage{subfigure}
\usepackage{amssymb}
\usepackage{multirow}
\usepackage{tabularx}
\usepackage{amsmath}
\usepackage{braket}
\usepackage{csquotes}
\usepackage{enumitem}
\graphicspath{{figs/}}
 \usepackage{lipsum}
\usepackage{mathrsfs}
\usepackage{MnSymbol}
\newcommand{\eps}[0]{\varepsilon}

\newcommand{\PyLIT}[0]{\texorpdfstring{P\MakeLowercase{y}LIT}{PyLIT} }

\newcommand{\beq}{\begin{equation}}
\newcommand{\eeq}{\end{equation}}
\newcommand{\bea}{\begin{eqnarray}}
\newcommand{\eea}{\end{eqnarray}}

\newcommand{\dom}{\,\mathrm{d}\omega}

\newcommand{\R}{\mathbb{R}}

\newcommand{\e}{\mathrm{e}}
\newcommand{\ba}{{\boldsymbol{\alpha}}}

\newcommand{\bF}{{\boldsymbol{F}}}

\newcommand{\mc}[1]{\mathcal{#1}}

\newcommand{\bs}[1]{\boldsymbol{#1}}

\usepackage{booktabs}
\usepackage{tikz}
\usepackage{pgfplots}
\pgfplotsset{compat=1.18} 
\usepackage{float}
\makeatletter
\let\newfloat\newfloat@ltx
\makeatother
\usepackage{graphicx,subfigure}
\usepackage{algorithm}
\usepackage{algpseudocode}
\usepackage{todonotes}

\usepackage[colorlinks=true,linkcolor=blue,urlcolor=blue,citecolor=blue]{hyperref}

\begin{document}
% ---------- %
% title page %
% ---------- %
%\title{Inverse two-sided Laplace transform in quantum many-body theory:\texorpdfstring{\\} generalized framework and numerical implementation\\\textcolor{red}{TC does not like this title, can it be changed?} \textcolor{green}{ABR: Yes, I also thought about that; my Ideas:}\\
%Reformulation of Analytic Continuation problem in quantum many-body theory\\
%Analytic Continuation in quantum many-body theory with kernel-based methods\\
%PyLIT: Reformulation and numerical implementation of Analytic Continuation problem due to kernel-based methods: \\
%}

\begin{frontmatter}

%% Title, authors and addresses

%% use the tnoteref command within \title for footnotes;
%% use the tnotetext command for theassociated footnote;
%% use the fnref command within \author or \affiliation for footnotes;
%% use the fntext command for theassociated footnote;
%% use the corref command within \author for corresponding author footnotes;
%% use the cortext command for theassociated footnote;
%% use the ead command for the email address,
%% and the form \ead[url] for the home page:
%% \title{Title\tnoteref{label1}}
%% \tnotetext[label1]{}
%% \author{Name\corref{cor1}\fnref{label2}}
%% \ead{email address}
%% \ead[url]{home page}
%% \fntext[label2]{}
%% \cortext[cor1]{}
%% \affiliation{organization={},
%%             addressline={},
%%             city={},
%%             postcode={},
%%             state={},
%%             country={}}
%% \fntext[label3]{}

\title{PyLIT: Reformulation and implementation of the analytic continuation problem using kernel representation methods}

\author[label1,label2,label3]{Alexander Benedix Robles}
\ead{a.benedix-robles@hzdr.de}

\author[label1,label2]{Phil-Alexander Hofmann}
\ead{p.hofmann@hzdr.de}

\author[label1,label2]{Thomas Chuna}
\ead{t.chuna@hzdr.de}

\author[label1,label2]{Tobias Dornheim}
\ead{t.dornheim@hzdr.de}

\author[label1,label2,label4]{Michael Hecht}
\ead{m.hecht@hzdr.de}

\affiliation[label1]{organization={Center for Advanced Systems Understanding (CASUS)}, 
                     postcode={D-02826}, 
                        city={Görlitz}, 
                        country={Germany}}
                        
\affiliation[label2]{organization={Helmholtz-Zentrum Dresden-Rossendorf (HZDR)}, 
                    postcode={D-01328},
                    city={Dresden}, 
                    country={Germany}}
                    
\affiliation[label3]{organization={Technische  Universität  Dresden},  postcode={D-01062},  city={Dresden},  country={Germany}}
\affiliation[label4]{organization={Mathematical Institute, University Wrocław}, postcode={50-384}, city={Wrocław}, country={Poland}}

%% use optional labels to link authors explicitly to addresses:
%% \author[label1,label2]{}
%% \affiliation[label1]{organization={},
%%             addressline={},
%%             city={},
%%             postcode={},
%%             state={},
%%             country={}}
%%
%% \affiliation[label2]{organization={},
%%             addressline={},
%%             city={},
%%             postcode={},
%%             state={},
%%             country={}}

%% Abstract
\begin{abstract}
Path integral Monte Carlo (PIMC) simulations are a cornerstone method for studying quantum many-body systems, such as warm dense matter and ultracold atoms. The analytic continuation needed to estimate dynamic quantities from these simulations amounts to an inverse Laplace transform, which is an ill-conditioned problem. If this challenging problem were surmounted, dynamical observables such as the dynamic structure factor (DSF) $S(q,\omega)$---a key property e.g.~in x-ray and neutron scattering experiments---could be extracted from the imaginary-time correlation functions $F(q,\tau)$ estimates. 

Although of fundamental importance, the analytic continuation problem remains challenging due to its ill-posedness, and state-of-the-art techniques continue to deliver unsatisfactory results. To address this challenge, we express the DSF as a linear combination of kernel functions with known Laplace transforms that have been tailored to satisfy its physical constraints, \textit{e.g.}, detailed balance. Then we employ least-squares optimization regularized with a Bayesian prior estimate to determine the coefficients of this linear combination.
We explore various regularization term, such as the commonly used entropic regularizer, as well as the uncommon Wasserstein distance and $L^2$-distance. We also explore techniques for setting the regularization weight. A key outcome and contribution is the open-source package PyLIT (\textbf{Py}thon \textbf{L}aplace \textbf{I}nverse \textbf{T}ransform), which leverages Numba for C-level performance, unifying the presented formulations. PyLIT's core functionality is kernel construction and optimization.

In our applications, we find that PyLIT's DSF estimates share many qualitative features with other more established methods. Drawing from our insights, we identify three key findings. Firstly, independent of the regularization choice, utilizing non-uniform grid point distributions reduced the number of unknowns and thus reduced our space of possible solutions. Secondly, the Wasserstein distance, a previously unexplored regularizer, performs as good as the entropic regularizer while benefiting from its linear gradient. Thirdly, future work can meaningfully combine regularized and stochastic optimization.
\end{abstract}

%% Keywords
\begin{keyword}
%% keywords here, in the form: keyword \sep keyword

%% PACS codes here, in the form: \PACS code \sep code

%% MSC codes here, in the form: \MSC code \sep code
%% or \MSC[2008] code \sep code (2000 is the default)

\end{keyword}

\end{frontmatter}

\section{Introduction}

%1.	Opening: The paragraph needs to establish what task I am trying to accomplish. It ought to start with a general statement that captures the reader's interest and introduces the broad topic or problem that my research addresses. The paragraph ought to provide some context and background information to highlight the importance and relevance of the study and end by stating the specific research question or objective of your paper.
Understanding the dynamics of interacting quantum many-body systems is an active frontier in physics, quantum chemistry, and related disciplines. For quantum many-body systems, path integral Monte Carlo (PIMC) methods provide formally exact simulations~\cite{cep,boninsegni2,Takahashi_Imada_PIMC_1984}, but are limited to the \emph{imaginary-time} domain. These simulations generate imaginary-time correlation functions (ITCF) and an analytic continuation (AC) is needed to determine dynamic quantities of interest ~\cite{boninsegni1,Filinov_PRA_2012,nolting, Krilov_JCP_1999,Rabani_PNAS_2002}. In the present work, we focus on the analytic continuation of imaginary-time density--density correlation functions $F(q,\tau)=\braket{\hat{n}(q,\tau)\hat{n}(-q,0)}$. Here $\hat{n}$ denotes the single-particle density operator in reciprocal space. $F(q,\tau)$ is the usual intermediate scattering function~\cite{sheffield2010plasma} $F(q,t)$ evaluated at the imaginary argument $t=-i\hbar\tau$, where $\tau\in[0,\beta]$ and $\beta=1/k_\textnormal{B}T$ is the inverse temperature. Reliable estimates of $F(q,\tau)$ are currently being used to inform experimental diagnostics~\cite{Dornheim_review,Dornheim_T_2022,Dornheim_MRE_2023,Dornheim_PTR_2023,Schoerner_PRE_2023,shi2025firstprinciplesanalysiswarmdense,Dornheim_SciRep_2024,Dornheim_Science_2024}. In this work, we analytically continue PIMC ITCFs to obstain estimates of the dynamic structure factor (DSF) $S(q,\omega)$, a key property in scattering experiments, e.g.~with ultracold atoms~\cite{Filinov_PRA_2012, Ferre_PRB_2016, Filinov_PRA_2016, Dornheim_SciRep_2022, Boninsegni1996, Saccani_Supersolid_PRL_2012} or warm dense quantum plasmas~\cite{Dornheim_review, siegfried_review, kraus_xrts, Tilo_Nature_2023, Fletcher2015, boehme2023evidence, Gregori_PRE_2003}. %Prominent examples for such problems include the Matsubara Green function $G_\textnormal{M}(q,\tau)$ that is connected to the dynamic single-particle spectrum $A(q,E)$~\cite{boninsegni1,Filinov_PRA_2012,nolting} and the velocity--velocity ITCF that is related to the dynamic diffusion coefficient~\cite{Krilov_JCP_1999,Rabani_PNAS_2002}. 

The ITCF and DSF are related by a Laplace transformation $\mathcal{L}$ over $\pm \infty$ (\textit{i.e.}, double-sided Laplace transform)
\begin{eqnarray}\label{eq:Laplace}
    F(q,\tau) = \mathcal{L}\left\{S(q,\omega)\right\}[\tau] = \int_{-\infty}^\infty \textnormal{d}\omega \, S(q,\omega) e^{-\tau \omega} \ .
\end{eqnarray}
Here we employ Hartree atomic units throughout where $\hbar=1$ so that the frequency $\omega$ has energy units. PIMC simulations produce an estimate $F(q,\tau)$ at discrete $\tau$ points. This leads practitioners to solve the discretized version of \eqref{eq:Laplace}
\begin{align}\label{eq:discrete_Laplace}
    \bF = \bs{A} \bs{S},
\end{align}
where for a single value of $q$, $N_\tau$ values of $\tau_i$, and $N_\omega$ values of $\omega_j$ we have vector $F_i = F(q, \tau_i)$, vector $S_j=S(q, \omega_j)$, and matrix $A_{i,j} = e^{ - \tau_i \omega_j}$. Typically, $\tau_i$ and $\omega_i$ have equidistant spacing~\cite{Asakawa2001MEM, JARRELL1996133, tripolt2019comparison, hansen2019extraction}. In summary, practitioners are finding an estimate to the continuous AC problem \eqref{eq:Laplace} by solving the discrete formulation \eqref{eq:discrete_Laplace}.

%2.	Background and literature review: Previous attempts [A,B,C] to accomplish [task] have encountered problems [X,Y,Z]. In this paragraph, provide a more detailed background of the research topic. Discuss the existing knowledge, theories, and previous studies related to your research question. Highlight the gaps or limitations in the existing literature that your study aims to address. This paragraph should demonstrate your familiarity with the subject area and showcase the novelty of your research.
Conventionally, this discrete formulation is difficult in two ways. Firstly, the number of desired $\omega$ points $N_\omega > N_\tau$ making \eqref{eq:discrete_Laplace} an ill-posed problem. 
Secondly, the singular values of $A$ decay towards zero exponentially fast~\cite{epstein2008badtruth, istratov1999} which makes \eqref{eq:discrete_Laplace} increasingly ill-conditioned. Thus, increasing $N_\tau$ does not improve the capability of the discretisation to capture the AC problem. As a result of these two difficulties, the analytic continuation problem has remained an open challenge for decades.

Due to the importance of the DSF and the difficulty of the AC problem, a plethora of methods have been suggested over the years. Most often the inversion is formulated as an optimization problem with a regularization term $r(\cdot)$ and regularization weight $\lambda$ as
\begin{equation}
    \label{eq:typical_min_problem}
    \underset{\bs{S}}{\min}  \, \, \frac{1}{2}\|\bF - \bs{A} \bs{S} \|_2^2  + \lambda r(\cdot).
\end{equation}
Essentially, most works differ from each other in their choice of $r(\cdot)$ and how they handle the unknown $\lambda$. We give the following examples of regularized optimization. The maximum entropy method (MEM) selects the Shannon-Jaynes entropy regularized optimization and averages solutions over a collection of $\lambda$ values~\cite{JARRELL1996133, Boninsegni_maximum_entropy, Silver_PRB_1990, chuna2025dual, chuna2025estimates} according to Gull's Bayesian posterior~\cite{gull1989MEMBayesianWeighting} or selects a single best $\lambda$ value using the $\chi^2$-kink algorithm \cite{KAUFMANN2023ana_cont}. The Bayesian reconstruction method selects the Burnier-Rothkopf entropy regularized optimization and also averages solutions over $\lambda$~\cite{BurnierRothkopf2013bayesianreconstruction}. The sparse modeling approach, by Otsuki \emph{et al.}~\cite{Otsuki_PRE_2017,Otsuki_JPSJ_2020}, selects the $L^1$ norm regularized optimization and selects a single $\lambda$ value. Various stochastic optimization methods sample according to the unregularized optimization problem and average over the large number of independent, noisy random solutions for $S(q,\omega)$ to achieve a regularization effect, see, e.g., Refs.~\cite{Filinov_PRA_2012, Ferre_PRB_2016, Filinov_PRA_2016, Dornheim_SciRep_2022, Mishchenko_PRB_2000,Vitali_PRB_2010, shu2015stochastic} and this has been shown to be a generalized form of entropic regularization~\cite{beach2004identifying, Fuchs_PRE_2010}. While entropic and stochastic approaches constitute the majority of approaches, there are other optimization methods such as Backus-Gilbert Method \cite{backus1968, backus1970, press2007numerical}, the Gaussian Backus-Gilbert Method~\cite{hansen2019extraction}, and artificial neural networks~\cite{Fournier_PRL_2020,Yoon_PRB_2018}. Further, other approaches reformulate the problem as approximating $S(q,\omega)$ by Pad{\'e} approximants \cite{press2007numerical, tripolt2019comparison}, or reformulate the problem as approximating $S(q,\omega)$ in terms of its frequency moments~\cite{Dornheim_moments_2023}, \textit{i.e.}, the Hamburger problem~\cite{tkachenko_book,Tkachenko_CPP_2018,Vorberger_PRL_2012, Filinov_PRB_2023}. 
There are open source implementations of some of these methods in the $\text{ana}\_\text{cont}$ code \cite{KAUFMANN2023ana_cont} and the ACflow code \cite{huang2023acflow}.
Still, no general method of choice has emerged yet~\cite{tripolt2019comparison, Loon_PRB_2016, Goulko_PRB_2017}. %and t

% 3.	Objectives and research approach: Here, we alter approach B so that it doesn’t have instability [X].  Clearly state the specific objectives or goals of your study. Outline the main hypotheses or research questions that your research aims to answer. Describe the methodology or approach you have taken to investigate these questions. Briefly mention the experimental setup, theoretical framework, or computational models used in your research.
In this work, we generalize the typical approach \eqref{eq:typical_min_problem} so that the aforementioned difficulties can be addressed. This is done by expressing $S(q,\omega)$ as a linear combination of kernel functions with known Laplace transforms. By comparison, solving the typical formulation determines the coefficients of the linear combination of Dirac deltas and solving our optimization problem determines the coefficients of the linear combination of kernel functions. While representing a function via a set of kernels is common practice in the mathematical literature (\textit{e.g.} Galerkin methods in PDEs \cite{langtangen2019functionapproximation}), to the authors knowledge discussion of this has been neglected for the AC problem (except for in the trivial, Dirac-delta, limit that yields \eqref{eq:discrete_Laplace}~\cite{sandvik_PRB_1998}). Our generalization, makes the kernel set an input into the optimization problem and we use gradient descent to find the best coefficients for a linear combination of kernel functions (\textit{i.e.} Gaussian or Uniform kernels).

% 4.	Discuss the imact/applications of the solution you are providing. Discuss the potential applications, practical implications, or theoretical advancements that your study can contribute to the field. Emphasize the relevance and potential impact of your research to capture the readers' attention and highlight the significance of your work. With respect to the solution,
We publish a high-performance Python code base that implements these ideas: the \textbf{Py}thon \textbf{L}aplace \textbf{I}nverse \textbf{T}ransform (PyLIT) package, which is freely available in an \href{https://github.com/phil-hofmann/pylit}{online code repository}. Specifically, PyLIT contains a variety of kernel bases and regularization terms, which can be combined freely. This facilitates an easy comparison between different ans\"atze, thus giving a way to explore the inherent uncertainty in a solution of the considered AC problem. A key strength of our implementation is its high performance, which makes it easy to combine a selected AC solver with other applications or uncertainty quantification schemes. As an application, we consider the uniform electron gas (UEG) (i.e., the quantum version of the classical one-component plasma~\cite{Ott2018}) at warm dense matter conditions~\cite{review}, which has been the focus of numerous works over the last decade~\cite{roadmap}.

%5.	Overview of the paper: Provide a brief overview of the structure and organization of the paper. Mention the main sections or chapters and briefly describe what each section covers. This paragraph serves as a roadmap for the reader, helping them understand the flow of information in your paper.
The remainder of this paper is organized as follows: In Section \ref{sec:theory}, we present our formulation of the problem, describe how certain physical constraints are enforced on the solution, enumerate the regularization techniques which create a unique solution to our optimization problem, and describe our approach to determine the regularization weight. In Section \ref{sec:results}, we discuss the PyLIT interface and investigate the inversion of both synthetic and authentic ITCFs for the uniform electron gas. Finally, in Section \ref{sec:summary}, we draw conclusions from our results. %

\section{Theory of PyLIT\label{sec:theory}}

\subsection{Problem Formulation \label{sec:formulations}}
%We incorporate kernel functions whose double sided Laplace transform are known analytically, hence minimizing the numerical error. In this context, the commonly used formulation \eqref{eq:typical_min_problem} can be seen as an analogue of an uniform kernel function representation. To begin with, we represent the DSF as a linear combination of such kernel functions 
PyLIT solves a more general inversion problem than the typical \eqref{eq:typical_min_problem}. We formulate this problem by returning to the continuous Laplace transform equation \eqref{eq:Laplace} and representing the DSF as a linear combination of kernel functions
\begin{equation}\label{eq:linearcombo_of_kernels}
    S_\ba(\omega) = \sum_{j=1}^{m} \ba_j K_j(\omega).
\end{equation}
Here the $j$ index indicates all the parameters that determine the $j^\text{th}$ kernel. For instance, if $K_j(\omega)$ denotes a Dirac delta then $j$ denotes the location of the peak $\delta(\omega - \omega_j)$, as is typically the case in the stochastic optimization literature~\cite{sandvik_PRB_1998}. However, there are many kernels that can be used. For a Gaussian kernel, the index $j$ indicates the center $\mu_j$ and standard deviation $\sigma_j$. Upon inserting \eqref{eq:linearcombo_of_kernels} into the continuous Laplace equation \eqref{eq:Laplace}, we obtain the discrete inverse problem with regression matrix $\bs{R}$ that PyLIT solves
\begin{subequations} \label{eq:discrete_Laplace_linear_combo}
\begin{align} 
    \bF &= \bs{R} \ba \, ,
    \\ \bs{R}_{i,j} &= \int_{-\infty}^\infty \e^{-\bs{\tau}_i \, \omega} K_j(\omega) \dom = \mathcal{L}\{K_j(\omega)\}[\bs{\tau}_i].
\end{align}
\end{subequations}

To solve \eqref{eq:discrete_Laplace_linear_combo} for $\ba$, we formulate the regularized least square problem as
\begin{equation}
    \label{eq:min_problem}
    \underset{\ba}{\min} \, \, \frac{1}{2}\|\bF - \bs{R} \ba\|_2^2 + \lambda r(\cdot).
\end{equation}
The minimization of \eqref{eq:min_problem} yields the coefficients $\bm{\alpha}$ to approximate the DSF via \eqref{eq:linearcombo_of_kernels}. The choice of kernels directly affects the singular values of $\bs{R}$; this is verified in~\ref{app:automatickernelselection}. Thus, this formulation is preferred since the problem conditioning is adjustable. However, there is still no guarantee that \eqref{eq:min_problem} is well-posed and well-conditioned, thus the regularizer $r(\cdot)$ is included to provide this. In short, \eqref{eq:min_problem} extends \eqref{eq:typical_min_problem} because the inversion matrix no longer must be $e^{ - \tau_i \omega_j}$, which arises only in the limiting case that $\bs{S}$ is represented by a linear combination of Dirac deltas.

%, we observed in our experiments that $\bm{R}$ allows to have quantitatively fewer singular values close to zero than the typical choice $\bm{A}$. Specifically, selecting kernels in a stochastic manner, for instance by simulated annealing or greedy algorithms, induces a regularization effect without an explicit penalty term \cite{Hastie2008, Temlyakov2008}. We verify this effect in~\ref{app:automatickernelselection} by highlighting how the choice of kernels affects the singular values of $\bs{R}$. % Consequently, if the default (prior) model already provides a sufficiently accurate estimate of the true function, one can dispense with regularization and solve the generalized optimization for $\lambda \to 0$. From the optimization point of view, our formulation uses stochastic sampling to improve the optimization conditioning and constitutes an attempt to directly address the ill-posedness of the AC problem.

\subsection{Kernel \texorpdfstring{$K_j(\omega)$}  ~ Choices \label{sec:KernelChoices}}
Our regularized least squares problem \eqref{eq:min_problem} is determined by the choice of kernel functions $K_j(\omega)$. We use kernel functions whose Laplace transforms exist and are analytically known, thus eliminating the need for numerical integration. We emphasize the connection between the double sided Laplace transform of a probability distribution function (PDF) and the moment generating (MGF), as given by
\begin{align}
    \mathcal{L}\{S\}[\tau] = \text{MGF}(-\tau),    
\end{align}
providing access to literature in statistics. In this work, we consider the Uniform and Gaussian PDFs and their corresponding double sided Laplace transforms. An overview of these PDFs and their associated MGFs is given in Table \ref{tab:pdf-mgf}.
\begin{table}
\caption{\label{tab:pdf-mgf}Tabulation of the probability distribution functions (PDFs) $\widetilde{K}_j$ and their Laplace transforms $\mathcal{L}\{\widetilde{K}_j(\omega)\}[\tau]$ that are used to construct the kernels $K_j$ via \eqref{eq:constructed_kernel} for use in equations \eqref{eq:linearcombo_of_kernels}, \eqref{eq:discrete_Laplace_linear_combo}, and \eqref{eq:min_problem}. The constructed kernels $K_j$ fulfill \ref{C1}--\ref{C4}. However, the exponential decay PDF, marked with $^*$, is not considered in this work because it does not satisfy constraint \ref{C4} for all values of $b_j$, this is discussed more in \ref{C4}.}
\centering
\begin{tabular}{lll}
\toprule
    \textrm{Model} & $\widetilde{K}_j(\omega)$& $\mathcal{L}\{\widetilde{K}_j(\omega)\}[\tau]$\\
    \midrule\\[-1ex]
    Uniform & \(\displaystyle \frac{\mathbf{1}_{[\omega_j, \omega_{j+1}]}}{\omega_{j+1} -\omega_j}  \) & \(\displaystyle \frac{\e^{-\tau \omega_{j+1}} - \e^{-\tau \omega_j}}{-\tau(\omega_{j+1} - \omega_j)}\) \\[3ex]
    Gaussian & $\displaystyle   \frac{1}{\sigma_j \sqrt{2\pi}} \exp \left( {-\dfrac{1}{2} \left( \dfrac{\omega - \mu_j}{\sigma_j}\right)^2} \right)$ & $\exp\left( -\tau \mu_j + \frac{1}{2} \sigma_j^2 \tau^2 \right)$ \\[3ex]
    Laplace* & $\displaystyle \frac{1}{2b_j} \exp\left( - \frac{|\omega- \mu_j|}{b_j} \right)$ & $\displaystyle \frac{\exp(-\tau \mu_j)}{1+b_j^2\tau^2} $\\[3ex]
%    Cauchy* & \(\displaystyle \frac{1}{\pi} \left( \frac{\gamma}{(\omega - \omega_j)^2+\gamma^2} \right) \) & \(\exp(-\omega_j \tau - \gamma |\tau|)\) \\
\bottomrule
\end{tabular}
\end{table}

% While these MGFs are analytic expressions, they do not guarantee that a linear combination will satisfy the relevant conditions known to the quantum many-body community. We wish to estimate $S_\ba(\omega)$ and by manufacturing appropriate kernels and selecting appropriate optimizers we can produce solutions which satisfy important constraints.
There is no guarantee that a linear combination of Laplace transformed Uniform or Gaussian kernels will satisfy physical constraints known to the quantum many-body community. We lay out the relevant constraints in \ref{C1}-\ref{C4} and establish how to manufacture kernels $K_j$ from PDFs $\widetilde{K}_j$ that satisfy these constraints.
\begin{enumerate}[left=0pt,label={(C\arabic*)}]
    \item \label{C1} Consistency: The ITCF must converge to a sum of exponentials $F(\tau) \to \sum_j  c_j e^{- \tau \omega_j}$ in the limit, which is equivalent to require the kernels to converge to Dirac delta functions $\widetilde{K}_j \to \delta(\omega - \omega_j)$. Consistency is a priori fulfilled by our choice of kernels in Table~\ref{tab:pdf-mgf}.
    % Each of the kernels $\widetilde{K}_j(\omega)$ in Table~\ref{tab:pdf-mgf} reduce to a Dirac in some limit and therefore their MGF's reduce to $e^{- \tau \omega}$. %Since, $S_\ba$ is a linear combination then it is consistent.
    
    \item \label{C2} Nonnegativity: $S_{\bs{\alpha}}(\omega) \geq 0$. This condition is enforced by two separate choices. Firstly, the MGFs presented in Table \ref{tab:regularization_with_default} are non-negative. Secondly, the optimizer is constrained to nonnegative coefficients. Together, these choices ensure that the solution is a linear combination of nonnegative functions with nonnegative coefficients, which must be nonnegative.

    \item \label{C3} Detailed balance: $S_{\bs{\alpha}}(-\omega) = \e^{-\beta \omega} S_{\bs{\alpha}}(\omega)$~\cite{GiulianiVignale2008quantumtheory}. Detailed balance is handled by constructing kernels from the MGFs. Observe that the expression
    \begin{equation}\label{eq:constructed_kernel}
        K_j(\omega) \coloneqq \e^{\beta \omega} \widetilde{K}_j(-\omega) + \widetilde{K}_j(\omega)
    \end{equation}
    satisfies detailed balance. Further, observe that the Laplace transform of \eqref{eq:constructed_kernel} demonstrates the expected periodicity~\cite{Dornheim_T_2022,Dornheim_MRE_2023}
    \begin{equation}\label{eq:LaplaceTransformedKernel}
        \mc{L}\set{K_j}[\tau] = \mc{L}\set{\widetilde{K}_j}[\beta-\tau]+ \mc{L}\set{\widetilde{K}_j}[\tau] = \mc{L}\set{K_j}[\beta - \tau].
    \end{equation}
    
    \item \label{C4} Asymptotic dominance: 
    $\frac{\exp(-\omega \beta)}{S_\ba(\omega)} \to \infty \quad \text{for} \quad \omega \to \infty$. 
    This condition arises from the $\e^{\beta \omega} \widetilde{K}_j(-\omega)$ term in \eqref{eq:constructed_kernel}. If $\exp(\omega \beta)$ increase faster to $+\infty$  than $\widetilde{K}_j$ decreases to $0$ the function is bounded at $+\infty$. This constraint precludes the exponential PDF, shown in Table\ref{tab:pdf-mgf}, because if the PDF's chosen decay rate is slower than $\beta$ (\textit{i.e.}, the largest $\tau$ value ) the Laplace transform is unbounded.
\end{enumerate}
%of these linear map in terms of $\ba$, yielding the explicit formula for its Laplace transform
%\begin{equation}
%    F_\ba(\tau)  = \sum_{j=1}^{m} \ba_j \left(\mc{L}\set{\widetilde{K}_j}[\beta-\tau] + \mc{L}\set{\widetilde{K}_j}[\tau]\right).
%\end{equation}
%Thus, \ref{C3} is satisfied.

In summary, the kernels $K_j(\omega)$ that are used in equations \eqref{eq:linearcombo_of_kernels}, \eqref{eq:discrete_Laplace_linear_combo}, and \eqref{eq:min_problem} are constructed according to \eqref{eq:constructed_kernel}. By doing so (and optimizing with a nonnegativity constraint) our solutions will satisfy constraints~\ref{C1}--\ref{C4}. In our numeric implementation, the regression matrix~$\bs R$, from \eqref{eq:min_problem}, is constructed directly from $\mc{L}\set{\widetilde{K}_j}[\beta-\tau]+ \mc{L}\set{\widetilde{K}_j}[\tau]$ using the MGF formulas in Table \ref{tab:pdf-mgf}. The remaining challenge is selecting the hyperparameters for each of the $j$ kernels $\widetilde{K}_j(\omega)$.%; this is discussed in Section~\ref{app:automatickernelselection}.

\subsection{Regularizer \texorpdfstring{$r(\cdot)$} ~ Choices\label{sec:regularization}}
To define \eqref{eq:min_problem}, the regularization term $r(\cdot)$ must be chosen and the selection process requires thought. On a fundamental level, regularization terms specify a unique solution to an under-determined problem and reduce sensitivity to noise. This second effect is discussed precisely within the machine learning literature as the ``bias-variance trade-off''. Compared to the least squares estimate, which has no bias, the regularization term will produce an estimate that contains bias. The hope is that the variance of the estimator has been lowered sufficiently so that the total mean square error (MSE) is lower than that of the least-squares estimate \cite{james2013statisticallearning}. It is known that entropic regularization does not introduce spurious correlations~\cite{JARRELL1996133}. However, PyLIT's reformulation of the AC problem is the first of its kind and the kernels themselves introduce a bias variance trade-off. Thus, there are no existing studies establishing a best regularizer. As such, we have implemented many different regularization choices. 

There are many types of regularization terms in the literature. We give three examples: regularization that only depends on the solution $S_\ba$ (\textit{i.e.}, $r(S_\ba)$), regularization that only depends on the kernel coefficients $\ba$ (\textit{i.e.}, $r(\ba)$), and regularization that depends on the solution and a default model $D$ (\textit{i.e.}, $r(S_\ba, D)$. In this work, we consider three regularization terms that depend on a default model (\textit{i.e.} Wasserstein distance, entropy, and $L^2$-distance). The Wasserstein distance relies on the cumulative distribution function (CDF), which is defined as 
\begin{equation}\label{eq:CDF}
    \mathrm{CDF}[S_\ba](\omega) = \int_{-\infty}^\omega S_\ba(\omega') \, \text{d} \omega'. 
%    \mathrm{CDF}[S_\ba]: L^1(\mathbb{R}) \rightarrow L^1_{\mathrm{loc}}(\mathbb{R}), \quad S_\ba \mapsto \int_{-\infty}^x S_\ba(y) \dy. 
\end{equation} 
These regularization terms and their gradients are formulated in Table~\ref{tab:regularization_with_default}. 

\begin{table}
\centering
\caption{\label{tab:regularization_with_default}Regularization terms that use default model $D$. The entropy regularization penalizes deviations in $\log(S_\ba)$ from $\log(D)$ weighted by $\lambda$. \cite{kubo1957statistical, JARRELL1996133, chuna2025dual} The Wasserstein distance reglularization penalizes deviations in the accumulation in $S_\ba$ from accumulation in $D$ via their cumulative distribution functions (CDF) defined in \eqref{eq:CDF} \cite{yang2023wasserstein}. The $L^2$-distance regularization uniformly penalizes deviations in $S_\ba$ from $D$. The numerical implementation of these terms is given in Table~\ref{tab:NNN_implmentation}.}
\begin{tabular}{lll}
    \toprule
    regularizer & $r(S_\ba,D)$ & \( \nabla_{S_\ba} r(S_\ba, D) \) \\
    \midrule\\[-1ex]
     entropy & \(\displaystyle - \int_{-\infty}^\infty S_\ba(\omega) \ln \left( \frac{S_\ba(\omega)}{D(\omega)} \right) \dom \) & \( \displaystyle - \left( \ln\left(\frac{S_\ba(\omega)}{D(\omega)}\right) +1\right)\) \\[3ex]
     Wasserstein & \( \displaystyle \frac{1}{2}\left\|\operatorname{CDF}[S_\ba-D]\right\|_{L^2(\R)}^2 \) & $\displaystyle \operatorname{CDF}^* \operatorname{CDF} [S_\ba-D]$ \\[3ex]
     $L^2$-distance & $\displaystyle \frac{1}{2}\|S_\ba-D\|_{L^2(\R)}^2$ & $S_\ba-D$ \\
    \bottomrule		
\end{tabular}\label{tab:regularizers_analytic}
\end{table}

All three regularization terms considered in Table~\ref{tab:regularization_with_default} include a default model and have a unique minimum (\textit{i.e.} are convex). For the $L^2$-distance and Wasserstein distance their convexity is established by the quadratic nature of both optimization terms; for the entropy this is established elsewhere \cite{Asakawa2001MEM, van2014renyi}. However, we do not expect each minimum to yield the same solution. Axiomatic arguments conclude that the entropic regularizer does not introduce correlation across $\omega$ and that the $L^2$-distance does~\cite{JARRELL1996133}. The authors are not aware of investigations into the bias of the Wasserstein distance. The advantage of using these regularization terms is that a good choice of $D(\omega)$ can greatly reduce the bias of the unique minimum. The obvious challenges are to supply such a $D(\omega)$ and to weight the regularization appropriately via $\lambda$. %However, using the Cauchy Schwarz inequality one can show that the bias of the Wasserstein distance is not zero and that $W_2^2(u,v) \leq \frac{1}{2}(b-a)^2 \|u-v\|_{L^2(a,b)}$, which implies that the bias in the Wasserstein distance is smaller than in the $L^2$-distance.

\section{Overview of \PyLIT code\label{sec:PyLIToverview}}
The core functionality of the PyLIT code is kernel construction and optimization. For a given default model, regularization term, and regularization weight, PyLIT realizes a collection of kernels, defines \eqref{eq:min_problem}, and solves this optimization problem. Then PyLIT uses these optimal coefficients to evaluate \eqref{eq:linearcombo_of_kernels}, producing an estimate of the DSF. %We provide a kernel selection algorithm which takes a default model and yields a set of kernels.

% Paragraph about the selection of kernels.
%With respect to kernel construction, 
PyLIT has built-in kernel construction. For a given fixed number $n$ kernels, either Gaussian or uniform, we use a simulated annealing (SA) algorithm to select the best parameters of the kernels according to how well they fit the default model. The cost of the SA algorithm is less than the cost of PyLIT optimization and a single PyLIT optimization is on the order of seconds. The SA optimization is typically terminated before convergence by the max number of iterations. This is done intentionally to prevent biasing the kernel choice too strongly to the default model, but this makes the initialization of kernels very important. We do the following. For the uniform distribution kernels, we interpolate the CDF of the default model's gradient to select the $\omega$ grid spacing. This causes the kernels to have the tightest grid spacing where the gradient magnitude is highest. For the Gaussian kernels, we interpolate the CDF of the default model to place the centers $\mu_j$ near where the default model has the greatest relative magnitude. All Gaussian's share a common $\sigma$ that is initialized to the geometric mean of the smallest and largest $\mu$ spacing. The details of the kernel selection are given in~\ref{app:automatickernelselection}

% Paragraph about optimizer/regularizer.
With respect to PyLIT's optimization functionality, since the regularization terms considered here (\textit{i.e.}, $L^2$-distance, entropy, and Wasserstein distance) lead to strictly convex problems having unique minimums then gradient descent optimizers are best. Further, per constraint \ref{C2}, PyLIT must use a nonnegative optimizer. We choose the gradient-descent nonnegative Nesterov (NNN) optimizer \cite{Nesterov2012}. Other non-negative gradient descent methods exist such as nonnegative least squares \cite{Bro1997} and nonnegative Adaptive Moment estimation (ADAM) \cite{Kingma14}. However, the projection-based nonnegative least squares does not work for the entropy because the gradient of this regularizer is not a linear equation that can be solved as $\nabla L(\theta) = 0$. Additionally, the Nesterov iteration scheme is known to converge faster than ADAM, in fact, it is the optimal iterative method for these problems, as shown by Nesterov \cite[Section 2.2 Optimal Methods]{nesterov2018lectures}. Note that optimization is performed on coefficients $\ba$, but regularization is performed with respect to the solution $S_\ba$. In practice, an evaluation matrix $E$ comprised of the kernel is used to express regularization in terms of the coefficients $\ba$. More details of our NNN implementation are given in~\ref{app:nonnegative_nesterov}.

%By default, \(\mathcal{T}\) is fixed with the definition of the model. In the case of the Gaussian model, this is defined via the sampling points (i.e. the expected values) and variances (the width of the function). With a given default model, we can simply calculate the variances and expected values and select the model parameters based on these. In principle, this works, but it is unclear whether, for example, too many or too few sampling points and variances are selected. For this reason, we introduce an adaptive approach. This adaptive approach attempts to achieve better results if no default model is available. The detail of this adaptive appproach are laid out in~\ref{app:Adaptive}

% Paragraph about additional features. 
%PyLIT also contains functionality that can improve the quality of results. Specifically, we implemented a normalization and scaling of the data. This feature is standard data practice and protects the user from overflow errors. In addition, we implemented singular value decomposition of the kernel. This feature allows the user to remove singular vectors from $\bs{R}$ which correspond to small singular values, controlling how sensitive the solution is to noise. However, in practice, kernel selection improved conditioning enough that we do not use this feature in our final results. See~\ref{app:tau_scaling} for details on scaling the data and see~\ref{app:svd} for details on the singular value decomposition.
PyLIT also contains functionality that can improve the quality of results. Specifically, we implemented a normalization and scaling of the data. This feature is standard data practice and protects the user from overflow errors. See~\ref{app:tau_scaling} for details on scaling the data.

%Paragraph about regularization weight selection
PyLIT does not contain functionality to select the regularization weight $\lambda$. From the literature, the two primary approaches are Gull's Bayesian posterior \cite{gull1989MEMBayesianWeighting}, which averages solutions over $\lambda$ values, and the $\chi^2$-kink algorithm~\cite{KAUFMANN2023ana_cont}, which selects a single best value. External to the PyLIT code, we explore both options in our applications. See~\ref{app:lambda_selection} for details and comparative plots.
%we find good agreement, similar to what is seen in the entropic regularization literature.
% in two ways. First, the solution does not need to be expressed as a linear combination of exponential decays and more appropriate functions for representing the solution can be selected. Second, 

\section{Application of \PyLIT code} \label{sec:results}
In this section, we apply the PyLIT code to both synthetic and authentic data. The process of applying PyLIT has three major steps: Kernel selection (documented in~\ref{app:automatickernelselection}), optimization (documented in~\ref{app:nonnegative_nesterov}), and regularization weight selection (documented in~\ref{app:lambda_selection}). The figures within these appendices have been generated from the synthetic data, making our process more transparent to the reader. Thus, only the relevant physics theory and final AC problem estimates are discussed here. 

%\textcolor{blue}{To support our statement with improved condition and its significance, we compare the results with $\lambda = 0$, i.e. a common least squares of our formulation.} \textcolor{red}{TC: could you add that? (Use $L^1$ or $L^2$ reg for testing and set $\lambda = 0$)}
%This would be great future work! Maybe call it "regularization free analytic continuation". we could explore the stochastic optimization of the basis set and then use regularization terms without default models. 

\subsection{Theory supporting the estimation of dynamic structure factors from imaginary time correlation functions}
There are many known approximations of the true DSF $S(q,\omega)$ that could inform $D(\omega)$ for a given wavenumber $q$. The DSF is related to the system's dynamic linear density response function $\chi(q,\omega)$ via the formally exact fluctuation dissipation theorem~\cite{GiulianiVignale2008quantumtheory,Dornheim_review},
\begin{align} \label{eq:FDT}
    S(q,\omega) =  - \frac{1}{ \pi n} \frac{\text{Im} \chi(q,\omega)}{1-e^{- \beta  \omega} }.
\end{align}
Here $q$ and $\omega$ are the wavenumber and frequency of a weak external harmonic perturbation and $n$ denotes the number density. Without loss of generality, the density response is often expressed as~\cite{GiulianiVignale2008quantumtheory, ichimaru2018plasmavol1, ichimaru1982stronglycoupledplasma, kugler1} 
\begin{align}\label{eq:susceptibility}
    \chi(q,\omega) = \frac{\chi^0(q,\omega)}{1 - v(q) \left[1 - G(q,\omega)\right] \chi^0(q,\omega)}.
\end{align}
Here $\chi^0(q,\omega)$ is the non-interacting finite temperature Fermi gas susceptibility, the denominator includes the mean-field description of the dynamic density response, and $G(q,\omega)$ is the dynamic local field correction, which encodes the full many-body interactions into the one-body mean field interaction~\cite{atwal2002fullyconserving, chuna2025conservative}. $G(q,\omega)$ is directly related to the Kohn-Sham exchange--correlation (XC) kernel as $K(q,\omega) = v(q) G(q,\omega)$. Common models of the DSF vary in how the many-body interactions are handled~\cite{chuna2025conservative, dornheim_ML, stolzmann, stls, stls2, Dornheim_PRB_ESA_2021, Dornheim_PRL_2020_ESA, dornheim_dynamic}. 

Setting $G(q,\omega)\equiv 0$ in \eqref{eq:susceptibility} corresponds to the widely used random phase approximation (RPA). While being straightforward to compute, the RPA fails to capture the red-shift and the dispersion relation's (\textit{i.e.} the omega location of $S(q,\omega)$'s max value) roton-type feature that have been reported for the warm dense UEG and the strongly coupled electron liquid, respectively~\cite{Filinov_PRB_2023, dornheim_dynamic, koskelo2023shortrange, Dornheim_Nature_2022, Takada_PRB_2016, Takada_PRL_2002}. The roton type feature is defined by a non-monotonic dispersion relation that is negative $\partial_q w(q) < 0$ for small $q$ and positive $\partial_q w(q) > 0$ for large $q$. Both of red-shift and roton-type feature are electronic XC effects, which start to become important at Wigner-Seitz radii $r_s\gtrsim1$. Thus the RPA approximation is expected to break down in the $r_s \ge 10$ cases we consider in this work. 

A substantially more accurate approximation is given by $G(q,\omega) = G(q,0)$, which shall be denoted as \emph{static approximation}~\cite{dornheim_dynamic} throughout the remainder of this work. Dornheim et al.~\cite{dornheim_dynamic,Dornheim_Nature_2022} have shown that the static approximation is capable of delivering highly accurate results for $r_s\lesssim 4$. Moreover, it qualitatively captures the dispersion relation's roton-type feature for $r_s \gtrsim 4$, though the depth of the static roton minimum has been observed to be underestimated~\cite{chuna2025estimates, Filinov_PRB_2023, dynamic_folgepaper, Dornheim_PRE_2020}. A particular strength of this ans\"atze is that it can be evaluated exclusively on the basis of the PIMC input data for $F(q,\tau)$, this reduces the external bias that the default model introduces into the AC problem. Chuna et al.~\cite{chuna2025estimates} have very recently explored both the RPA and the static approximation as default models for the typical problem formulation \eqref{eq:typical_min_problem} with synthetic and authentic PIMC data for $F(q,\tau)$ and found the static approximation yields less volatile results with smaller uncertainty than the RPA.

To evaluate the static approximation of the density response and the DSF from authentic PIMC data for $F(q,\tau)$, we invert (\ref{eq:susceptibility}) in the static limit of $\omega\to 0$ as
\begin{align}\label{eq:static-approx}
    G(q,\omega=0) &= 1 + \frac{1}{v(q)} \left(\frac{1}{\chi(q,\omega=0)} - \frac{1}{\chi^0(q,\omega=0)}\right),
\end{align}
where $\chi(q,\omega=0)$ is computed from the imaginary-time version of the fluctuation dissipation theorem
\begin{align}\label{eq:chi_q}
    \frac{\chi(q, \omega=0)}{n \beta} = - \frac{1}{\beta} \int^{\beta}_0 \mathrm{d} \tau F(q,\tau).
\end{align}
See Ref.~\cite{Dornheim_MRE_2023} for a sample derivation of the static approximation~\eqref{eq:static-approx} and the imaginary time fluctuation dissipation theorem~\eqref{eq:chi_q}.

%The RPA approximation, which completely neglects $G(q,\omega$), does not exhibit this feature. The static approximation dispersion relation is not perfect though. Across studies~\cite{chuna2025estimates, chuna2025dual, Filinov_PRB_2023, dynamic_folgepaper, Dornheim_PRE_2020}, the fully dynamic local field correction produces a deeper roton-like feature in the dispersion relation across $q$'s and an incipient double peak structure in the individual $q$ cross sections. 

\subsection{Synthetic results\label{sec:synthetic}}
Our application to synthetic data is conducted to explore the impact of our kernel substitution. We are interested in how the estimates depend on the magnitude of noise in the synthetic signal. We are also interested in how unexplored regularization terms affect the solution. We consider the UEG at a Wigner-Seitz radius of $r_s=10$ at the electronic Fermi temperature $\theta=1/\beta E_\textnormal{F}$ with $E_\textnormal{F}$ being the usual Fermi energy. This system is located at the margins of the strongly coupled electron liquid regime~\cite{dornheim_electron_liquid}, where the static approximation exhibits the aforementioned roton-type feature in its dispersion relation at intermediate wavenumbers, but the RPA does not.

To construct our synthetic data, we compute the static approximation $S(q,\omega)$ using \eqref{eq:susceptibility}, \eqref{eq:static-approx}, and \eqref{eq:chi_q} with the authentic data described in Table~\ref{tab:simulations}. Then we Laplace transform [cf.~\eqref{eq:Laplace}] the static approximation to arrive at a noiseless signal $\bF^0$ and add Gaussian noise as $F_i = F^0_i + \delta F_i$, where
\begin{align}
    \delta F_i = F_i \, \mathcal{N}(0, \, \sigma_0).
\end{align}
Notice that the noise level is proportional to the signal; this is justified by Hatano's error analysis of the structure of PIMC errors~\cite{hatano1994data}. We consider noise levels $\sigma_0 = 10^{-1}, 10^{-2}, 10^{-3}$ to match the authentic PIMC data noise level.  

%When applying PyLIT to synthetic data, we use either 50 Gaussian kernels or 50 uniform kernels, and we use the RPA $S(q,\omega)$ (\textit{i.e.}, $G(q,\omega)=0$) as the default model. Furthermore, we consider various regularization terms. To select the regularization weight, we use Gull's approach for Gaussian kernels and the $\chi^2$-kink approach for uniform kernels. We use different methods because the term in Gull's approach that informs the Gaussian distribution of acceptable $\lambda$ values becomes numerically unstable for uniform kernels. In general, this work tends toward Gull's approach because it has no \textit{ad hoc} parameters, yields a well defined uncertainty estimate, and is more responsive to the choice of regularizer. For more details see~\ref{app:lambda_selection}.

When applying PyLIT to synthetic data, we use either 50 Gaussian kernels or 50 uniform kernels. The choice of 50 kernels is motivated by the fact that the number of data points $N_\tau$ is smaller than the number of kernels $N_\mu$, but sufficiently may to give flexibility in the functions that can be represented. Furthermore, we consider various regularization terms. To select the regularization weight, we use Gull's approach for Gaussian kernels and the $\chi^2$-kink approach for uniform kernels. We use different methods because the term in Gull's approach that informs the Gaussian distribution of acceptable $\lambda$ values becomes numerically unstable for uniform kernels. In general, this work tends toward Gull's approach because it has no \textit{ad hoc} parameters, yields a well defined uncertainty estimate, and is more responsive to the choice of regularizer. For more details see~\ref{app:lambda_selection}.

\begin{figure}
    \centering
    \includegraphics[width=0.33\linewidth]{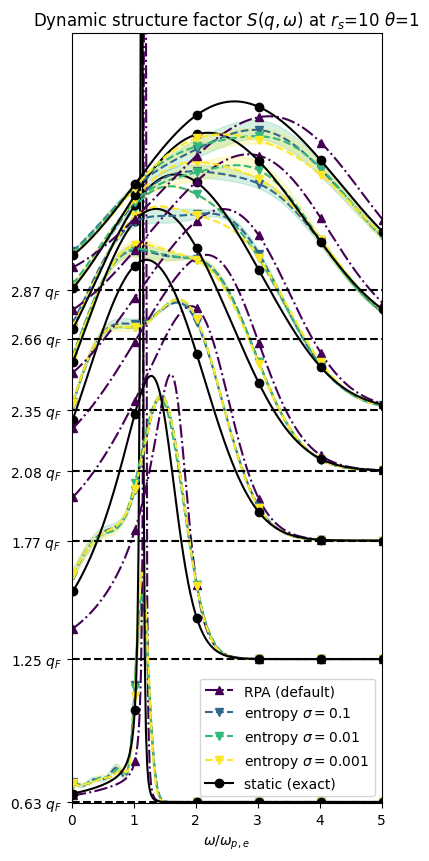}%
    \includegraphics[width=0.33\linewidth]{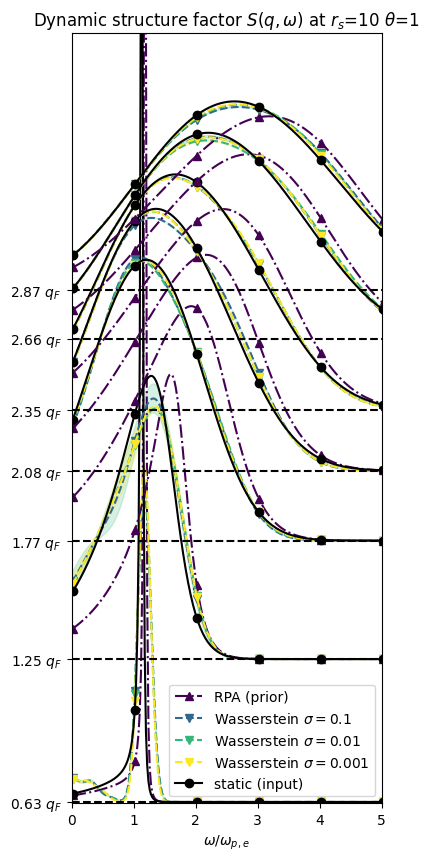}%
    \includegraphics[width=0.33\linewidth]{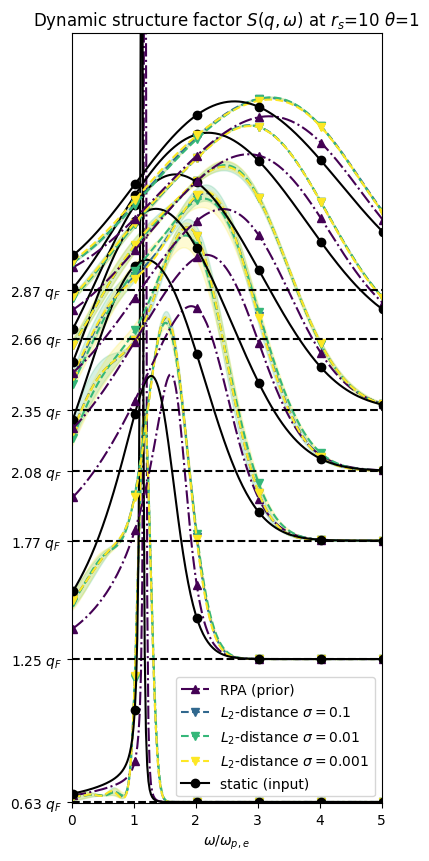}
    \caption{Each plot shows various PyLIT DSF estimates obtained with: 50 uniform kernels, the RPA default model, denoted (prior), and Gull's Bayesian averaging over $\lambda$~\cite{gull1989MEMBayesianWeighting}. The static approximation, denoted with (input), was used to construct the synthetic data and is the true solution. We compute the DSF at different values of the wavenumber normalized by the Fermi wavenumber $q/q_F$, as indicated by dash shadow line underneath each curve. The $\omega$ axis is normalized by the uniform electron gas's plasma frequency $\omega_{p,e}$. The $r_s$ and $\Theta$ values are indicated by the plot title. The regularization term is indicated by plot legend.; from left to right the cross entropy regularizer (Table~\ref{tab:regularizers_analytic}-entropy), the Wasserstein distance regularizer (Table~\ref{tab:regularizers_analytic}-WD), the $L^2$-distance regularizer (Table~\ref{tab:regularizers_analytic}-$L^2$-distance). Each plot contains different reconstructions of $S(q,\omega)$ obtained varying noise levels $\sigma_0 = 10^{-1}, 10^{-2}, 10^{-3}$.}
    \label{fig:synthetic_Gaussian_stacked-kslices}
\end{figure}

\begin{figure}
    \centering
    \includegraphics[width=0.33\linewidth]{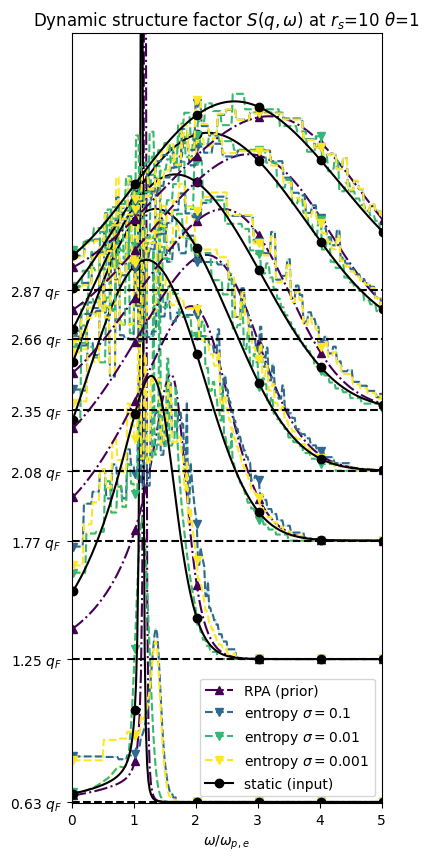}%
    \includegraphics[width=0.33\linewidth]{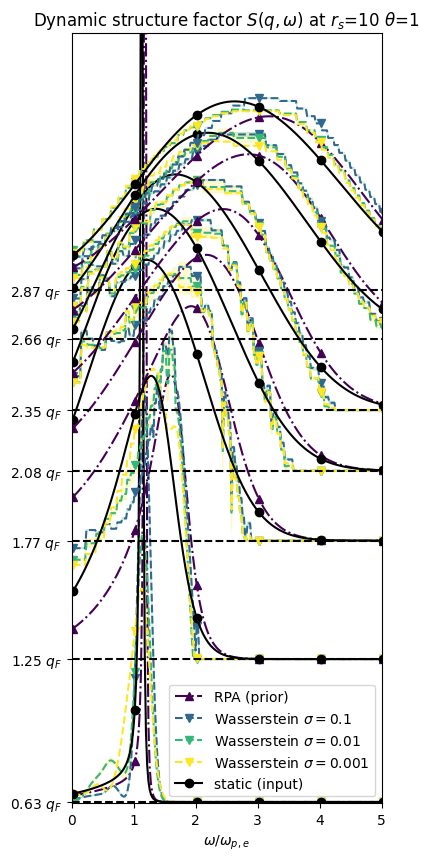}%
    \includegraphics[width=0.33\linewidth]{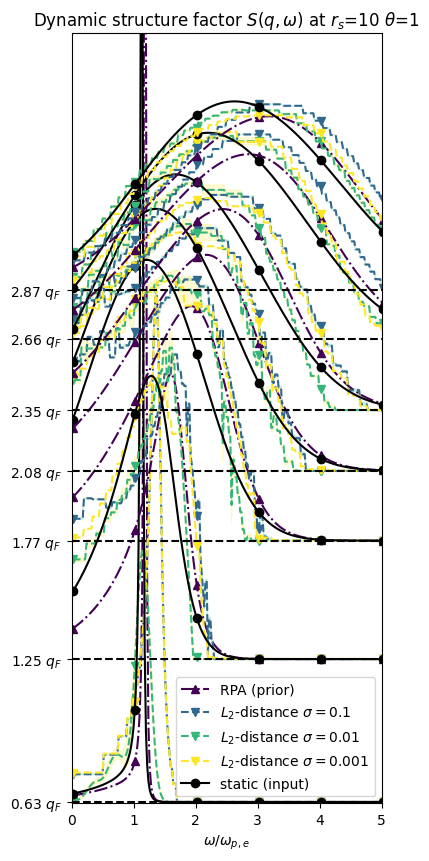}
    \caption{Each plot shows various PyLIT DSF estimates obtained with: 50 uniform kernels, the RPA default model, denoted (prior), and the $\chi^2$-kink regularization weight $\lambda$~\cite{KAUFMANN2023ana_cont}. The static approximation, denoted with (input), was used to construct the synthetic data and is the true solution. We compute the DSF at different values of the wavenumber normalized by the Fermi wavenumber $q/q_F$, as indicated by dash shadow line underneath each curve. The $\omega$ axis is normalized by the uniform electron gas's plasma frequency $\omega_{p,e}$. The $r_s$ and $\Theta$ values are indicated by the plot title. The regularization term is indicated by plot legend.; from left to right the entropy regularizer (Table~\ref{tab:regularizers_analytic}-entropy), the Wasserstein distance regularizer (Table~\ref{tab:regularizers_analytic}-WD), the $L^2$-distance regularizer (Table~\ref{tab:regularizers_analytic}-$L^2$-distance). Each plot contains different reconstructions of $S(q,\omega)$ obtained varying noise levels $\sigma_0 = 10^{-1}, 10^{-2}, 10^{-3}$.}
    \label{fig:synthetic_uniform_stacked-kslices}
\end{figure}

We plot the default model, the true solution, and reconstructions at various noise levels for Gaussian kernels in Fig.~\ref{fig:synthetic_Gaussian_stacked-kslices} and for uniform kernels in Fig.~\ref{fig:synthetic_uniform_stacked-kslices}. For the Gaussian kernels, the Wasserstein regularization matches the true signal the best and is robust to noise. The entropic regularizer does not match the true solution as well as Wasserstein, but as the noise decreases the estimate moves systematically towards the true solution. Comparatively, the $L^2$-distance does not match the true solution well and is barely affected by the noise level. This indicates that the $L^2$-distance severely over-regularizes. For the uniform kernels, none of the solutions perform particularly well. Though the entropic regularizer shares many qualitative features with the true solution for large $\omega$, its total variation increases drastically at small $\omega$, manifesting a volatile estimate. This behavior may be explained by arguments which establish that entropic regularization, unlike other regularizers, does not correlate values across $\omega$~\cite{JARRELL1996133}. %Hence the total bias introduced by the entropic regularizer will be less than either of the other regularization terms.


Comparing the two different kernel choices, we find a superior performance of the Gaussian compared to the uniform basis.
Empirically, the combination of the Gaussian kernels with the Wasserstein distance based regularization gives the best results, although the entropic regularization, too, predicts the expected roton-type dispersion due to the XC-induced red-shift of spectral weight in $S(q,\omega)$ compared to RPA.
We will thus focus on Wasserstein and entropic regularization throughout the remainder of this work.

%Comparing across both Gaussian and uniform kernels, we see the uniform kernels, as a whole, perform worse than Gaussian kernels. Regardless, the general trend that as the noise level diminishes the solutions tend towards the true solution exists across all approaches. Moving forward we will consider only Gaussian kernels. Overall, Gaussian kernels with Wasserstein distance is the best performing. Yet, both the entropic and Wasserstein regularization yield signals that peak at smaller $\omega$ values than the RPA approximation; thus, creating the expected roton feature. Based on this synthetic problem, we expect that Gaussian kernels with the entropic and Wasserstein regularization will yield the best results. Moving forward, these will be the results we focus on. 

%\subsubsection{residuals}
% paragraph discussing some plots which visualize the process for the reader.
%The choice of kernels determines which functions can be represented well. This resembles the basis selection problem in polynomial fitting, where fitting a polynomial of large order gives fit flexibility, but overfits the data and worsens the conditioning of the associated Vandermonde matrix. As in polynomial fitting, the consequence of well conditioned kernels is that the chi-sq cannot be made zero, nonetheless the residuals are numerically small. This is demonstrated in Fig.~\ref{fig:residuals}

\subsection{PIMC results\label{sec:PIMC_results}}
Here we present the PyLIT's estimates of $S(q,\omega)$ for the ITCF data described in Table~\ref{tab:simulations}. Our PIMC data contains $1000$ independent runs, generated via the open-source ISHTAR code~\cite{ISHTAR}, which is a canonical implementation~\cite{Dornheim_PRB_nk_2021} of the worm algorithm by Boninsegni et al.~\cite{boninsegni1, boninsegni2}. The error analysis on this data accounts for the fermion cancellation problem (\textit{i.e.} sign problem~\cite{dornheim_sign_problem,troyer}) using Hatano's error analysis~\cite{hatano1994data}. Further, we verified Hatano's formula, finding excellent numerical agreement with the leave-one-out binning error estimate~\cite{berg2004introduction}. This data set is the same one that was analyzed by Chuna et al.~\cite{chuna2025estimates}, regardless this application constitutes an original contribution on the structure of the finite temperature strongly coupled electron liquid because there is no known analytic form. In addition to the case of $r_s=10$ that has been analyzed on the basis of synthetic data in Sec.~\ref{sec:synthetic} above, here we also consider the more strongly coupled case of $r_s=100$. From the perspective of AC, this regime is particularly interesting, as we expect a more pronounced impact of the fully dynamic LFC $G(q,\omega)$ and, therefore, more pronounced deviations from the default model that corresponds to the static limit of $G(q,0)$.
\begin{table}
\caption{Tabulation of the physical parameters that describe our path integral Monte Carlo simulations. For Fermi energy $E_F = \hbar^2 k_F^2 / 2m_e$, temperature $T$, and Wigner-Seitz radius $r_s$, we list the quantum degeneracy parameter $\Theta = k_\textnormal{B} T / E_F$, the coupling parameter $\Gamma = \frac{1}{r_s} \left( (k_\text{B} T)^2 + E_F^2 \right)^{-1/2}$, the Fermi wavenumber $k_F$, the plasma frequency $\omega_{p,e}$, the inverse temperature $\beta$ in units of energy, the number of particles in the simulation, the volume of the simulation (\textit{i.e.} length of the periodic boundary cubed), the independent seeds (\textit{i.e.} the number of independent MCMC simulations), and the typical error in the ITCF values $\mathcal{O}[\delta F]$ obtained using Hatano's error analysis~\cite{hatano1994data}.}\label{tab:simulations}
\vspace{.1cm}
\centering
    \begin{tabular}{l l l}
        \toprule
            & $r_s=10$, $\theta=1$ & $r_s=100$, $\theta=1$ 
         \\ \hline $\Gamma$ & 7.68 & 76.8 
         \\ $k_F$ (1/Bohr) & $1.919 \times 10^{-1}$ & $1.919 \times 10^{-2}$
         \\ $\omega_{p,e}$ (Hartree) & $5.477 \times 10^{-2}$ & $1.732 \times 10^{-3}$
         \\ $\beta$ (Hartree) & $5.430 \times 10^{1}$ & $5.430 \times 10^{3}$ 
         \\ particles & 34 & 34
         \\ volume ($\text{cm}^3$) & $2.11 \times 10^{-20}$ & $2.11 \times 10^{-17}$  
         \\ ind. seeds & 1000 & 1000 
         \\ $\mathcal{O}[\delta F]$ & $10^{-4}$ & $10^{-4}$ \\
         \bottomrule
    \end{tabular}
\end{table}

\begin{figure}
    \centering
    \includegraphics[width=\linewidth]{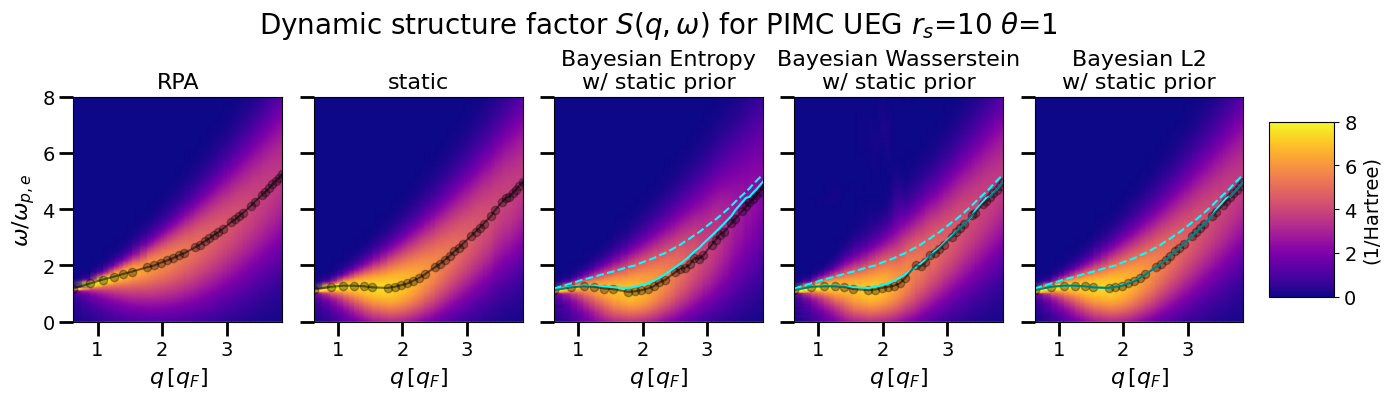}
    \includegraphics[width=\linewidth]{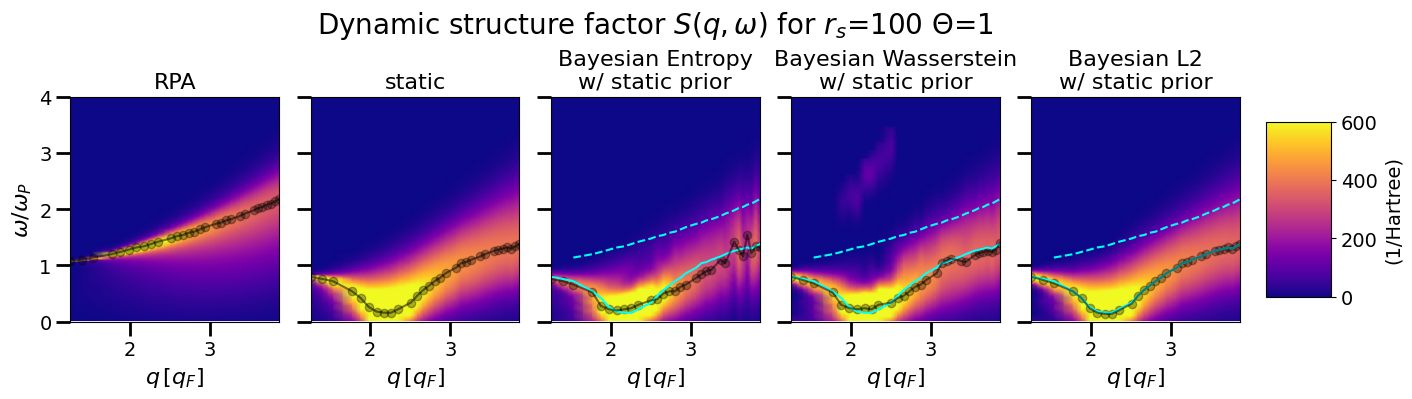}
    \caption{Heatmaps of the DSF $S(q,\omega)$ estimated by PyLIT. The $\omega$ axis is normalized by the plasma frequency $\omega_{p,e}$ and the $q$ values are normalized by the Fermi wavenumber $q_F$. The $r_s$ and $\Theta$ values indicated by the plot title, and the regularization weight procedure, regularization term, and default model are indicated by the plot subtitle. The dispersion relation for the RPA is plotted in all PyLIT estimates as a dashed cyan line and the dispersion relation for the static approximation is plotted as a solid cyan line.}
    \label{fig:heatmaps-authentic-Bayesian}
\end{figure}

In Fig.~\ref{fig:heatmaps-authentic-Bayesian}, we show heatmaps of the DSF for both $r_s=10$ (top row) and $r_s=100$ (bottom row) at the electronic Fermi temperature, $\theta=1$. 
The left column corresponds to the familiar RPA, which is accurate in the limit of small $q$ (long wavelength limit with an incipient delta peak at the electronic plasma frequency $\omega_{p,e}$) and large $q$ (short wavelength regime with the single-particle dispersion relation $\omega\sim q^2$), but does not capture the correct behavior when the wavelength $\lambda=2\pi/q$ is of the same order as the average interparticle distance, i.e., $q\sim2q_\textnormal{F}$, where $q_\textnormal{F}$ is the Fermi wavenumber~\cite{GiulianiVignale2008quantumtheory}. 
In contrast, the static approximation (second column from the left) exhibits a significant roton feature, which is particularly pronounced for $r_s=100$. 

\begin{figure}
    \centering
    \includegraphics[width=0.33\linewidth]{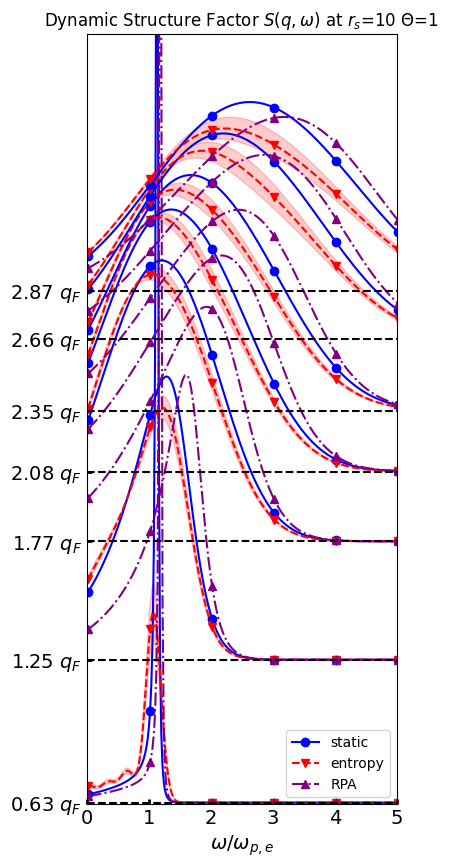}%
    \includegraphics[width=0.33\linewidth]{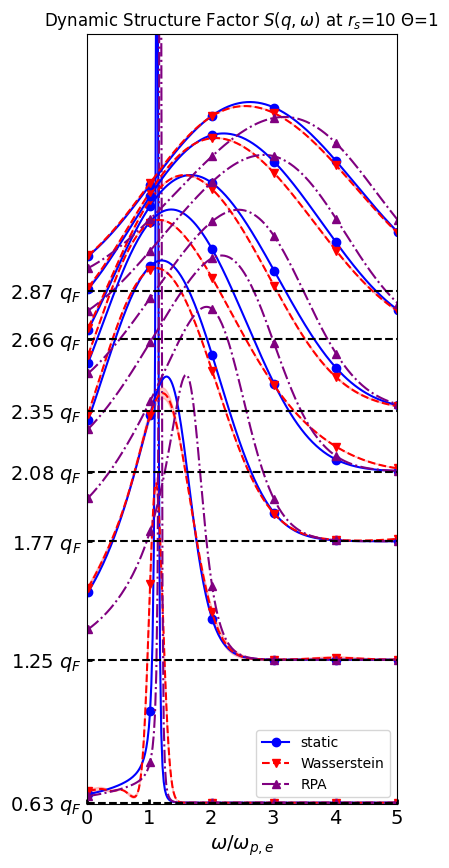}%
    \caption{Cross sections from the DSF heat maps plotted in Fig.~\ref{fig:heatmaps-authentic-Bayesian}. The cross sections are offset along the $y$-axis and the associated $q$ values are indicated by dash shadow line underneath each curve. The $r_s$ and $\Theta$ values indicated by the plot title, and regularization term indicated by plot legend. From left to right the entropy regularizer (Table~\ref{tab:regularizers_analytic}-CE)), the Wasserstein distance regularizer (Table~\ref{tab:regularizers_analytic}-WD), the $L^2$ distance regularizer (Table~\ref{tab:regularizers_analytic}-$L^2$)}
    %; this order matches DSF heatmaps from Fig.~\ref{fig:heatmaps-authentic-Bayesian}}
    \label{fig:stacked-kslices_rs10}
\end{figure}

\begin{figure}
    \centering
    \includegraphics[width=0.33\linewidth]{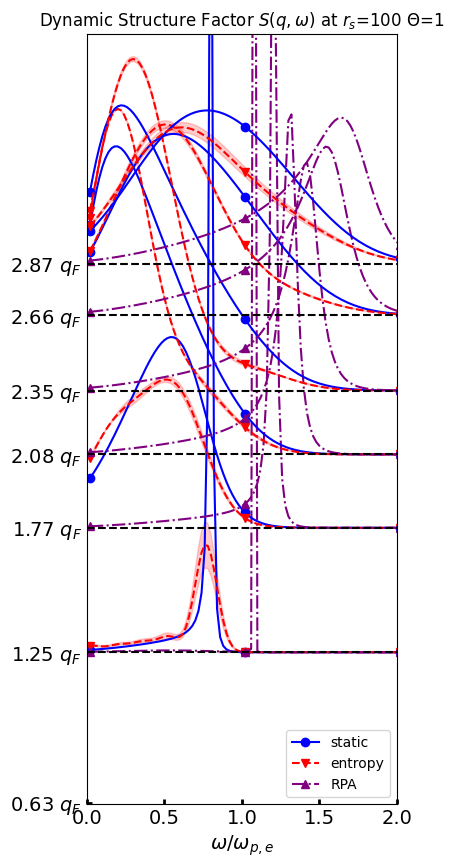}%
    \includegraphics[width=0.33\linewidth]{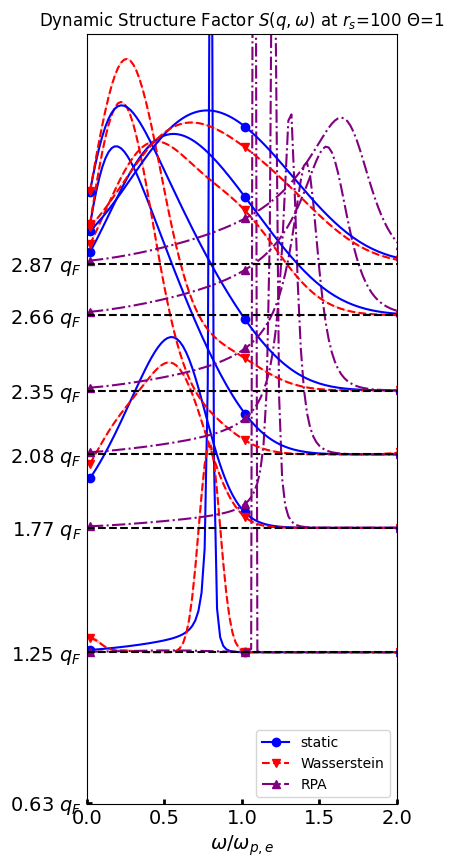}%
    \caption{Cross sections from the DSF heat maps, $q$ values indicated by dash shadow line underneath each curve, $r_s$ and $\Theta$ values indicated by the plot title, regularization term indicated by plot legend. From left to right the entropy regularizer (Table~\ref{tab:regularizers_analytic}-CE)), the Wasserstein distance regularizer (Table~\ref{tab:regularizers_analytic}-WD). We do not include $q=0.63 q_F$ in this plot because the width of the peak is Dirac-delta-like and cannot be resolved by our $\omega$ resolution.}
    \label{fig:stacked-kslices_rs100}
\end{figure}

Let us next turn to our new AC results, the final three columns. The third column from the right in Fig.~\ref{fig:heatmaps-authentic-Bayesian} shows AC results using the entropic regularizer, which overall resemble the Wasserstein regularizer, second from the right. Finally, the $L^2$ norm (right-most column) basically reproduces the default model and, thus, over-regularizes; this is consistent with our analysis of synthetic data shown in Sec.~\ref{sec:synthetic} above. Evidently, the results qualitatively follow the static approximation (solid cyan curve), but with a more pronounced roton minimum for $r_s=10$. This can be seen particularly well in Fig.~\ref{fig:stacked-kslices_rs10}, where we show the frequency dependence of $S(q,\omega)$ for selected wavenumbers for both regularizers at $r_s=10$. We note that the deeper roton minimum compared to the static approximation is consistent with previous investigations, see Refs.~\cite{chuna2025estimates, chuna2025dual, Filinov_PRB_2023, dynamic_folgepaper, Dornheim_PRE_2020}. In contrast, at $r_s=100$, both the entropic and Wasserstein regularization yield a more shallow minimum than the static approximation. This disagrees with the one other estimate at this $r_s$ value obtained by Chuna et al.~\cite{chuna2025estimates}. The reduced minimum can be seen particularly well in Fig.~\ref{fig:stacked-kslices_rs100}, where we show the frequency dependence of $S(q,\omega)$ for selected wavenumbers for both regularizers at $r_s=100$. We see here that the uncertainty that arises from varying the regularization weight $\lambda$ is small for both entropic and Wasserstein regularization. However, these error estimates do not include the uncertainty arising from the data, as quantified by leave-one-out binning, which Chuna et al.~\cite{chuna2025estimates} have shown to be as important as the error arising from the regularization weight. As such, PyLIT would need to be post-processed with resampling statistics to make more definitive conclusions. None-the-less it is interesting that PyLIT matches existing estimates at $r_s=10$, but predicts a diminished roton peak for $r_s=100$. 

We examine DSF heat maps cross section in more detail in Fig.~\ref{fig:stacked-kslices_rs10} and Fig.~\ref{fig:stacked-kslices_rs100}. These plots contain cross sections from the DSF obtained with entropic and Wasserstein regularization. We see that both estimates decrease the maximum DSF value for all $q$ in the case of $r_s=10$. Yet, for $r_s=100$ the maximum DSF value decreases for large $q$, but increases for intermediate $q$. Both of these behaviors, at $r_s=10$ and $r_s=100$ have been seen in DSF estimates obtained with the maximum entropy method~\cite{chuna2025estimates}.

Finally, we examine the differences that exist between the DSF obtained via entropy and Wasserstein regularizers. Using the entropic regularizer yields larger error bars than the Wasserstein distance for both $r_s=10$ and $r_s=100$ and in the $r_s=100$ case, using the entropic regularization yields more erratic dispersion relation. These observations are particularly interesting given that, in~\ref{app:lambda_selection} for the synthetic data, the $\chi^2$ behavior was consistent across all regularization terms, but Gull's Bayesian posterior selected larger regularization weights for the entropic regularizer. Using these results to make sense of our authentic results, this indicates that Gull's Gaussian approximation term and the regularization term are accounting for the increased variability in the entropic solutions and balancing this by selecting more regularized solutions. By comparison, the Wasserstein distance produces spurious peaks that can be faintly seen in Fig.~\ref{fig:heatmaps-authentic-Bayesian}. These peaks exist because the Bayesian posterior is selecting regularization weights that are much smaller than for the entropic regularizer.

\section{Summary and Discussion\label{sec:summary}}
In this work, we have generalized the typical regularized optimization problem which defines analytic continuation. This reformulation relies on representing the DSF as a linear combination of kernel functions with analytically known Laplace transforms. We recover the typical problem formulation in the limit that our kernels are chosen to be Dirac delta functions on a uniform grid. By comparison, solving the typical formulation determines the coefficients of the linear combination of Dirac deltas and solving our new optimization problem determines the coefficients of the linear combination of kernel functions. We demonstrate that the kernels directly impact the ill-posedness of the inversion in~\ref{app:automatickernelselection}. Thus, this reformulation constitutes an attempt to address the AC problem's conditioning in a mathematically consistent way.

We have gained algorithmic insights from the new formulation. Firstly, we have shown that the choice of kernel functions affects the conditioning of the inversion problem. Without well chosen kernels, the results could be even worse than what the typical formulation would provide. This insight also provides a way to improve the typical formulation. While there are no kernel widths/standard deviations, there are grid points, we recommend exploring grid point distributions that are not uniform to reduce the number of unknowns and thus shrink the solution space. Secondly, we have shown that linear combinations of Gaussian distributions can represent the solution to the analytic continuation. Currently, regularized optimization approaches weight many Dirac deltas to best fit the data. This insight also provides a way to improve stochastic sampling approaches. stochastic optimization typically constructs a set of Dirac deltas that fit the data~\cite{sandvik_PRB_1998,beach2004identifying, Fuchs_PRE_2010, Sandvik_PRE_2016} and yet the approach does not need to be limited to Dirac deltas. We recommend stochastic optimization algorithms explore linear combinations of Gaussian distributions. Thirdly, in~\ref{app:automatickernelselection} we describe how PyLIT selects the kernel set, which exposes a meaningful connection between regularized and stochastic optimization. We stochastically sampled kernel hyperparameters, but fit the kernel coefficients to our default model. From the perspective of stochastic optimization, this reduced the dimensionality of the search space. Future investigations may wish to merge stochastic and regularized optimization into a single algorithm, varying the number of kernels as well.
%by sampling kernel sets, using gradient descent to set the kernel coefficients, and then accepting or rejecting the kernel set by comparing against the data rather than the default model. %To do this, the PyLIT code would be nested withinin a simulated annealing algorithm that would allow the number of kernels vary and assess the collection of kernels by their ability to match the data rather than default model.

Our major result is a comprehensive and user-friendly Python implementation of the ideas presented in Section~\ref{sec:theory} ``Theory of PyLIT''. Our implementation helps users to readily explore our new formulation. We call the library PyLIT (Python Laplace Inverse Transform) [\href{https://github.com/phil-hofmann/pylit}{online code repository}] and we have detailed the implementation and functionality of PyLIT within appendices~\ref{app:automatickernelselection}, \ref{app:nonnegative_nesterov}and \ref{app:tau_scaling}. PyLIT uses stochastic optimization to tune the hyperparameters of either uniform or Gaussian kernels to a given default model and then conducts regularized optimization with a non-negative Nesterov optimizer to fit the kernel coefficients to a signal. We explore entropic, Wasserstein, or $L^2$-distance regularization options. For further details, such as sample notebooks, job scripts, or experiment configurations, please refer to the instructions and descriptions in the GitRepo. A key strength of PyLIT is its implementation, which allows for a straightforward integration into other applications and sampling schemes. Here, we use PyLIT in combination with different approaches to selecting a regularization parameter $\lambda$, the different approaches are discussed in~\ref{app:lambda_selection}.
%PyLIT does not contain functionality to select the regularization weight $\lambda$ and the 

We remind the reader that, to the best of our knowledge, neither the Wasserstein nor the $L^2$-distance regularizations have been explored in the analytic continuation literature. To address these unknowns, we have compared different approaches for setting the regularization weight on different regularizers. In~\ref{app:lambda_selection}, we compared the Wasserstein distance and the $L^2$-distance to the entropic regularizer, using either Gull's Bayesian posterior~\cite{gull1989MEMBayesianWeighting} or Kaufmann and Held's $\chi^2$-kink algorithm~\cite{KAUFMANN2023ana_cont}. We find that the $\chi^2$-kink selects a very similar regularization weight across regularizations and, in contrast, Gull's Bayesian posterior selects different regularization weight across regularizations. We take the Bayesian posterior's responsiveness to the choice of regularization as an advantage. For either approach, the synthetic data application demonstrated that as noise decreased the solution matched the input signal better. In this work, we have tended toward the Bayesian posterior. However, we find no conclusive evidence that one is better than the other. Future work ought to establish a rigorous error estimate for the $\chi^2$-kink algorithm's regularization selection. Overall we find for the synthetic example, the combination of the Wasserstein regularization with the Gaussian kernel functions has yielded the best results.

From a physical perspective, we have further substantiated the claims of a roton-like feature existing in the finite temperature electron liquid. For $r_s=10$, we estimate a deeper roton feature than the static LFC, as it is expected by many investigations~\cite{dornheim_dynamic,Dornheim_Nature_2022,chuna2025estimates,Filinov_PRB_2023}. However, at larger $r_s$, we estimate a shallower roton feature than the static approximation; this disagrees with the one other estimate at this $r_s$ value~\cite{chuna2025estimates}. However, without a more robust uncertainty quantification procedure (\textit{e.g.}, leave-one-out binning), which accounts for uncertainty arising from the data, we refrain from drawing conclusions. %Finally, see in the Wasserstein regularizer estimate no evidence of a double peak structure, but in the entropic regularization we do see an incipient double peak structure. As this is the very first application of Wasserstein distance regularizer to analytic continuation, we refrain from drawing conclusions from this observation.

PyLIT can be of immediate and future use for a gamut of practical applications. First, PyLIT can be directly applied to the analytic continuation of existing and upcoming PIMC datasets e.g.~for a variety of ultracold atom systems~\cite{Boninsegni1996,Boninsegni_maximum_entropy,Ferre_PRB_2016,Filinov_PRA_2012,Filinov_PRA_2016,Dornheim_SciRep_2022}. In this regard, a particularly enticing possibility is the consideration of warm dense quantum plasmas where both electrons and nuclei are treated dynamically on the same level~\cite{Dornheim_MRE_2024,Dornheim_Science_2024}. In addition to being interesting in their own right, such investigations will be useful for the diagnostics of extreme states of matter via x-ray Thomson scattering measurements~\cite{Tilo_Nature_2023,Dornheim_Science_2024,dornheim2024modelfreerayleighweightxray,schwalbe2025density} and might give important new insights about the accuracy of previously used approximate models for the DSF~\cite{bellenbaum2025estimatingionizationstatescontinuum,moldabekov2025applyingliouvillelanczosmethodtimedependent}.
Second, we intend to generalize PyLIT to other AC problems, such as the reconstruction of the single-particle spectral function $A(q,\omega)$ from the Matsubara Green function $G_\textnormal{M}(q,\tau)$~\cite{boninsegni1,Filinov_PRA_2012,PhysRevB.97.115164}. Third, the numerical inversion of a two-sided Laplace transform, in principle, allows for the deconvolution e.g.~of an experimental x-ray scattering signal to separate the influence of the combined source-and-instrument function from the actual dynamic structure factor~\cite{Gawne_JAP_2024}.
Finally, we note that possibility to perform a multi-dimensional analytic continuation of higher-order imaginary-time correlation functions~\cite{Dornheim_JCP_ITCF_2021} to compute e.g.~dynamic three-body structure factors and dynamic non-linear density response functions~\cite{Dornheim_PRR_2021,Dornheim_review,vorberger2024greensfunctionperspectivenonlinear} on a true \emph{ab initio} level.

\section*{Acknowledgments}
ABR, PAH and TC contributed equally to this work.
TC acknowledges Tom Gawne for insightful discussions on HPC and node-to-node communication.
This work was partially supported by the Center for Advanced Systems Understanding (CASUS), financed by Germany’s Federal Ministry of Education and Research (BMBF) and the Saxon state government out of the State budget approved by the Saxon State Parliament. This work has received funding from the European Research Council (ERC) under the European Union’s Horizon 2022 research and innovation programme
(Grant agreement No. 101076233, "PREXTREME"). 
Views and opinions expressed are however those of the authors only and do not necessarily reflect those of the European Union or the European Research Council Executive Agency. Neither the European Union nor the granting authority can be held responsible for them.
This work has received funding from the European Union's Just Transition Fund (JTF) within the project \emph{R\"ontgenlaser-Optimierung der Laserfusion} (ROLF), contract number 5086999001, co-financed by the Saxon state government out of the State budget approved by the Saxon State Parliament.
Computations were performed on a Bull Cluster at the Center for Information Services and High-Performance Computing (ZIH) at Technische Universit\"at Dresden, and at the Norddeutscher Verbund f\"ur Hoch- und H\"ochstleistungsrechnen (HLRN) under grant mvp00024. %, and on the HoreKa supercomputer funded by the Ministry of Science, Research and the Arts Baden-W\"urttemberg and
by the Federal Ministry of Education and Research.
%\end{acknowledgements}

\appendix
\section{Automatic kernel selection with default model \label{app:automatickernelselection}}
Selecting a finite number of kernels to represent an unknown function is, at best, naive. In general appropriate kernel selection is a large and active field, sometimes referenced as dictionary learning. We do not attempt a full investigation of this field. We limit ourselves in this investigation, which is primarily focused on the problem formulation and we take as a given that practitioners have a model of the expected solution that is reasonably good. For example, recent investigations by Chuna et al.~\cite{chuna2025estimates} have shown this is the case for the data we consider in this work. 

Assuming a default model that qualitatively reflects the true solution, we want a set of hyperparameters that can represent this model well. Although a relatively small number of kernels may suffice to approximate the default model (indeed, even two or three Gaussian kernels could adequately capture the default model), using as many as 50 kernels ensures that the representation maintains sufficient flexibility to accommodate deviations from the default model in realistic scenarios. We employ a simulated annealing (SA) optimization algorithm to select the parameters for these 50 kernels. The SA algorithm adds exponential noise to keep the kernel parameters positive and halts when the collection of kernels can fit the default model better than a given tolerance. The pseudo code is given as below. In practice we give the simulated annealing algorithm a very short 1000 iterations to explore the search space. This gives the ability to do some optimization, but prevents the algorithm from overfitting to the default model.

\begin{algorithm}
\caption{Simulated annealing}\label{alg:SA}
\begin{algorithmic}[1]
\Require Support points: $\boldsymbol{\omega} \in (\mathbb{R^+})^{N_\omega}$, target values: $\bs{D} \in (\mathbb{R^+})^{N_\omega}$, initial kernel hyperparameters: $\{\text{old\_parameters}\}$, goodness-of-fit tolerance: $\varepsilon_{SA} > 0$, annealing beta: $\beta_{SA} \in \mathbb{R}^+$, annealing step size: $\eta_{SA} \in \mathbb{R}^+$ 

\For{$i = 1$ to $\text{max\_iter}$}
    \State $\#$ Propose new kernel hyperparamters  
    %\State $\sigma' = \sigma \, \exp( \eta_{SA} \, \mathcal{N}(0,1) )$
    %\For{$j=0$ to $\text{num\_kernels}$}
    %    \State $\mu_j' = \mu_j\exp( \eta_{SA} \,\mathcal{N}(0,1) )$
    %\EndFor
    \State $\text{new\_parameters} = \text{old\_parameters} \times \exp( \eta_{SA} \, \mathcal{N}(0,1) )$
    \State
    \State $\#$ Find best approximation of the default model with new parameters.
    \State Collect new kernels into an evaluation matrix $E \in \mathbb{R}^{n_\omega \times \text{num\_kernels}}$.
    \State Solve $E \, \bs{\alpha} = \bs{D}$ for $\bs{\alpha}$.
    \State Store $\varepsilon_i = \Vert E \, \bs{\alpha} - \bs{D} \Vert_2^2$
    \State
    \State $\#$ Probabilistic Acceptance Step
    \If{ $(\varepsilon_i - \varepsilon_{i-1} ) < 0$ or $\mathcal{U}(0,1) < \exp(- \beta_{SA}(\varepsilon_i -\varepsilon_{i-1} ))$}
        \State $\#$ Check if update satisfies tolerance.
        \If{$\varepsilon_i < \eps_{SA}$}
            \State $\text{best\_parameters} = \text{new\_parameters}$
        \EndIf
        \State
        \State $\#$ Check if update is global optima and cool SA temperature.
        \If{$\varepsilon_i < \varepsilon_{i-1}, ... \varepsilon_0$}
            \State $\text{best\_parameters} = \text{new\_parameters}$
            \State $\beta_{SA} = \gamma_{SA} \, \beta_{SA}$
            \State $\eta_{SA} = \gamma_{SA} \, \eta_{SA}$
        \EndIf
    \EndIf
\EndFor
\State return best\_parameters
\end{algorithmic}
\end{algorithm}

\subsection{For Gaussian PDF}

We discuss here particulars of the SA optimization algorithm for Gaussian kernels, including implementation of the kernels, how the hyperparameters are initialized, and the effects of hyperparameter tuning. From Table~\ref{tab:pdf-mgf} and \eqref{eq:LaplaceTransformedKernel}, we present the Laplace transformed Gaussian kernel
\begin{align}
\mathcal{L}\{\widetilde{K}_j(\omega)\}[\tau] = \exp\left( -\tau \mu_j + \frac{1}{2} \sigma_j^2 \tau^2 \right)\ .
\end{align}
The regression matrix to be inverted is given as
\begin{align}
    R_{i,j} = \exp\left( -(\beta-\tau_i) \mu_j + \frac{1}{2} \sigma_j^2 (\beta-\tau_i)^2 \right) + \exp\left( -\tau_i \mu_j + \frac{1}{2} \sigma_j^2 \tau_i^2 \right)\ .
\end{align}
To restrict the optimization search space we enforce a shared value across all kernels $\sigma =  \sigma_j$. Thus, the SA algorithm varies the centers $\{ \mu_j \}$ and a single shared $\sigma$.

The initialization of $\{\mu_j\}, \, \sigma$ is crucial. Emperically, we have found good results by uniformly sampling the default model's CDF, placing $\{\mu_j\}$ according to where the default model has the most importance. Then selecting $\sigma$ to be the geometric mean of the largest and smallest centroid spacing $\Delta\mu_i =  \mu_j - \mu_{j+1}$ (\textit{i.e.} $\sqrt{\Delta\mu_\text{max} \Delta\mu_\text{min}}$). The CDF sampling process is visualized for two different default models below in Fig.~\ref{fig:mu-intialization}.
\begin{figure}
    \centering
    \includegraphics[width=0.5\linewidth]{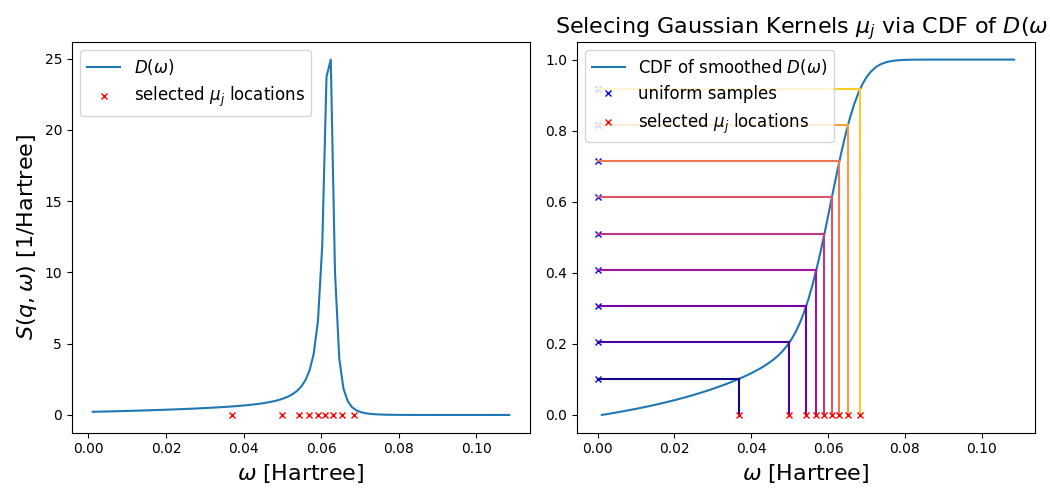}%
    \includegraphics[width=0.5\linewidth]{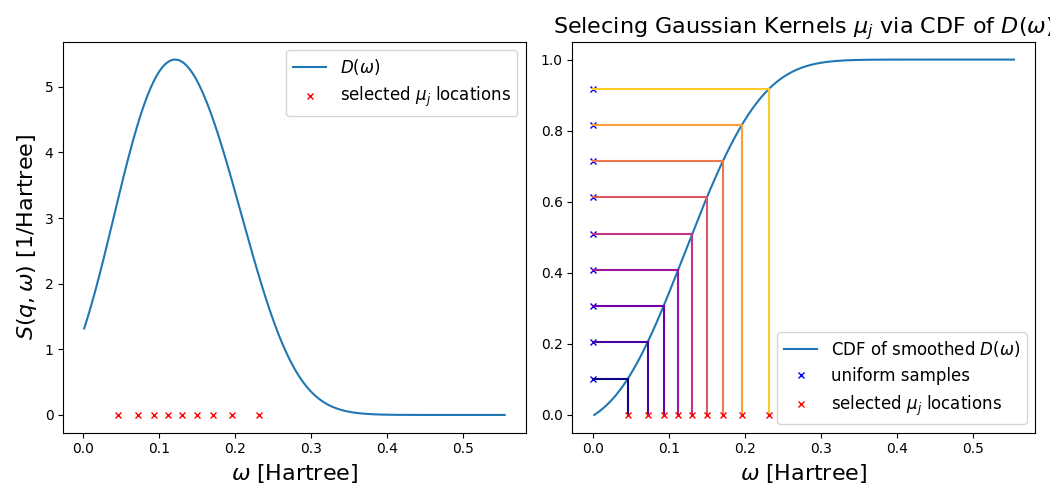}
    \caption{Plot visualizing how the Gaussian kernels centroid locations $\{ \mu_j \}$ are initialized for the simulated annealing algorithm (only a subset of the 9 of the 50 centroids are visualized here). Left: plot visualizing the centroids computed for a sharply peaked default model (\textit{i.e.} RPA $r_s=10$, $\Theta=1$, $q = 0.63 q_F$). Right: plot visualizing the centroids computed for a smoother default model (\textit{i.e.} RPA $r_s=10$, $\Theta=1$, $q = 2.66 q_F$)}
    \label{fig:mu-intialization}
\end{figure}

\begin{figure}
    \centering
    \includegraphics[width=0.5\linewidth]{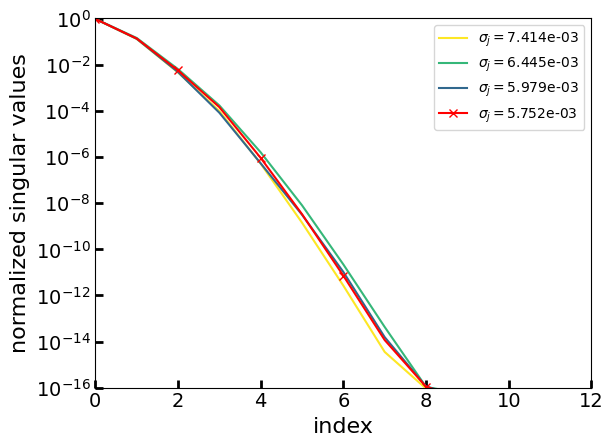}%
    \includegraphics[width=0.5\linewidth]{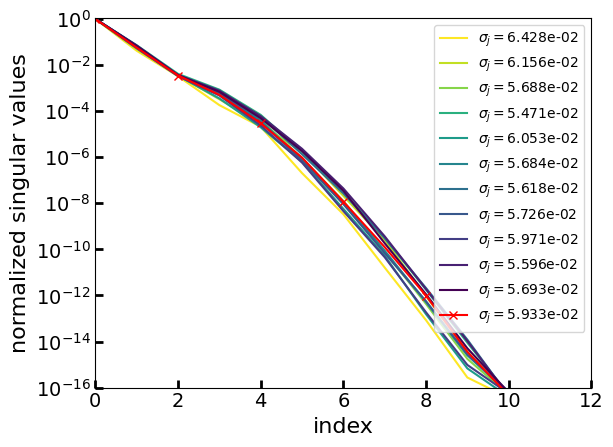}
    \caption{Plot of the singular values associated to Gaussian kernel regression matrix $R$ from \eqref{eq:min_problem} for each accepted update of the simulated annealing algorithm. The curves are plotted in acceptance order so the coloration visualizes the simulated annealing algorithm. Left: simulated annealing algorithm conducted on sharply peaked default model (\textit{i.e.} RPA $r_s=10$, $\Theta=1$, $q = 0.627 q_F$). Right: simulated annealing algorithm conducted on smooth default model (\textit{i.e.} RPA $r_s=10$, $\Theta=1$, $q = 2.66 q_F$). Comparatively, the regression matrix on the right with further spaced $\{ \mu_i\}$ and wider $\sigma$ has more non-zero singular values than the regression matrix on the left. This establishes that, even if the problem is ill-posed, the hyperparameters impact the conditioning. Furthermore, these plots demonstrate that best match of the default model (indicated with red colors and marked with x's) does not necessarily correspond to the best or worst conditioning. This establishes that, matching the data well does not mandate the problem must be ill conditioned.}
    \label{fig:conditioningViaSingularValues}
\end{figure}

As claimed in the main text, the choice of the kernels directly affects the conditioning of the problem, which is understood in terms of the singular values of the regularization matrix. The more singular values that are close to zero, the more sensitive the inversion is to noise as there are more flat search directions in the least squares optimization problem. We observe how our hyper parameter sampling alters the singular values of regression matrix $R$ from \eqref{eq:min_problem} in Fig.~\ref{fig:conditioningViaSingularValues}.

\begin{figure}
    \centering
    \includegraphics[width=0.5\linewidth]{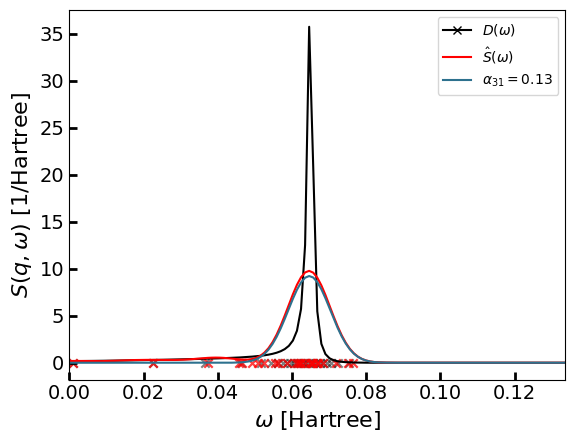}%
    \includegraphics[width=0.5\linewidth]{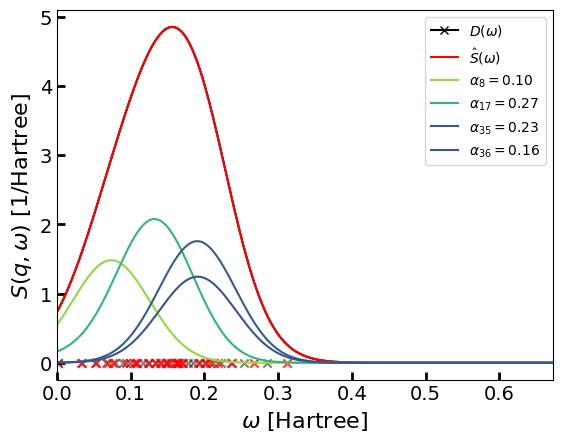}
    \caption{Plot of the final estimate of the default model obtained by the simulated annealing algorithm for Gaussian kernels. Left: RPA default model with $r_s=10$, $\Theta=1$, $q = 0.627 q_F$. Right: RPA default model with $r_s=10$, $\Theta=1$, $q = 2.66 q_F$.}
    \label{fig:SA_Gaussian_reconstructions}
\end{figure}

For completeness, we also plot the simulated annealing algorithm's final reconstruction in Fig.~\ref{fig:SA_Gaussian_reconstructions}. When considering Fig.~\ref{fig:SA_Gaussian_reconstructions} Left, it is important to keep in mind that the SA algorithm is given little opportunity to optimize, this leads to widths that are too broad to perfectly represent the peak. Allowing the SA is to converge leads to very narrow Dirac-delta like Gaussian distributions that are clustered near the RPA peak. Such a tight clustering of narrow Gaussian distributions reduces the space of solutions that can be represented, leading to poor final estimates.  

\subsection{For uniform PDF}
We discuss here particulars of the SA optimization algorithm for Gaussian kernels, including implementation of the kernels, how the hyperparameters are initialized, and the effects of hyperparameter tuning. From Table~\ref{tab:pdf-mgf} and \eqref{eq:LaplaceTransformedKernel}, we present the Laplace transformed uniform kernel 
%As initially mentioned, we choose probability densities as kernel functions, $\|K_j\|_{L^2(\R)}=1$, $ j =1,\dots,m$. Consequently, the Laplace transform $ \mathcal{L}\{K_j\}$ is given by the \emph{moment-generating-function}. 
\begin{equation*}
   \mathcal{L}\{K_j(\omega)\} = \frac{\e^{-\tau \omega_{k+1}} - \e^{-\tau \omega_k}}{-\tau(\omega_{k+1} - \omega_k)} + \frac{\e^{-(\beta - \tau) \omega_{k+1}} - \e^{-(\beta - \tau) \omega_k}}{-(\beta - \tau) (\omega_{k+1} - \omega_k)}
\end{equation*} 
However, the denominator makes this expression numerically unstable in the $\tau \rightarrow 0$ or $(\beta - \tau) \rightarrow 0$. but the value of the kernel is known in these limits. 
\begin{equation*}
   \lim_{\tau \to 0} \frac{\e^{-\tau \omega_{k+1}} - \e^{-\tau \omega_k}}{-\tau(\omega_{k+1} - \omega_k)} = -1
\end{equation*} 
Thus, we avoid this issue by expanding the Laplace transform of the uniform kernel in a Taylor series at $\tau = 0$: 
\begin{equation*}
        \mathcal{L}\left\{ \frac{\mathbf{1}_{[\omega_j, \omega_{j+1}]}}{\omega_{j+1}-\omega_j} \right\}[\tau] = \sum_{j=k}^{\infty} (-1)^k \frac{\tau^{k-1}}{k!} = -1 + \frac{1}{2}\tau(\omega_{j+1}+\omega_j) +\cdots\,, 
\end{equation*}
enabling stable evaluation. 

Uniform kernels best approximate flat functions. To this end, we compute the gradient of the default model $\partial_\omega D(\omega)$ and tightly place the kernels where the functions value changes most rapidly. We average nearest neighbors in the gradient because the numerical gradient is known to be sensitive and protect against overfitting to the default model's gradient. From this smeared gradient, we compute the CDF and then uniformly sample the CDF. The gradient CDF sampling process is visualized for two different default models below in Fig.~\ref{fig:grid-intialization}.
\begin{figure}
    \centering
    \includegraphics[width=0.5\linewidth]{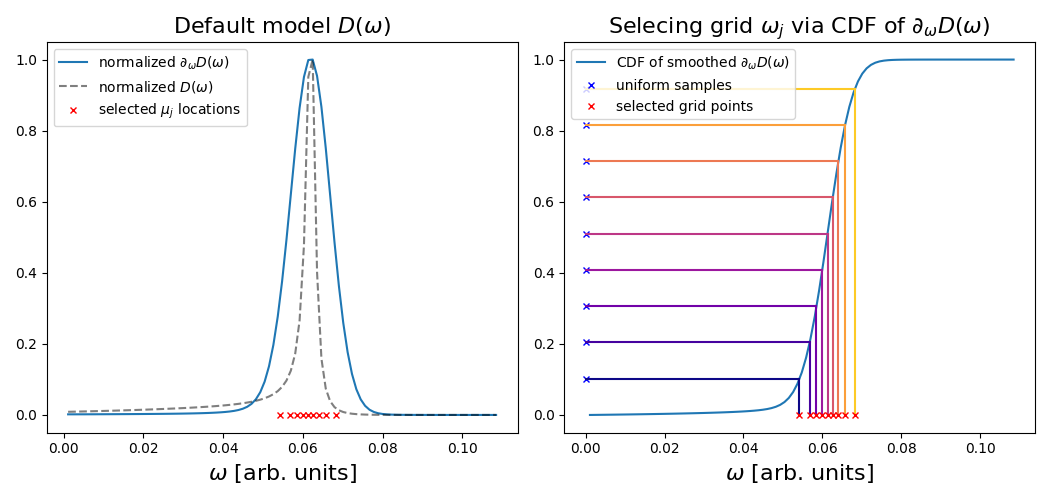}%
    \includegraphics[width=0.5\linewidth]{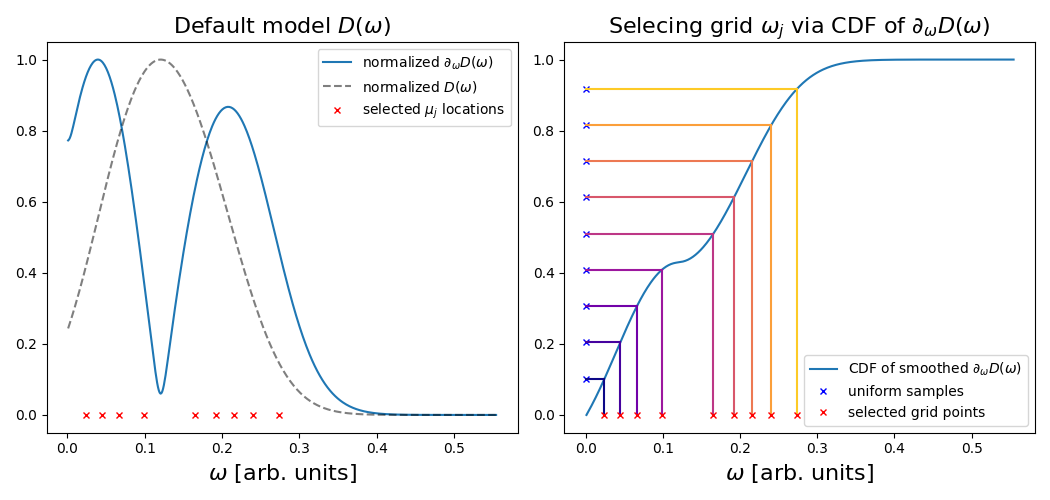}
    \caption{Plot visualizing how the $\{ \omega_j \}$ grid for uniform kernels are initialized for the simulated annealing algorithm (only a subset of the 9 of the 50 grid points are visualized here). Left: plot visualizing $\{ \omega_j \}$ grid for a sharply peaked default model (\textit{i.e.} static approximation $r_s=10$, $\Theta=1$, $q = 0.627 q_F$). Right: plot visualizing the $\{ \omega_j \}$ grid for a smoother default model (\textit{i.e.} static approximation $r_s=10$, $\Theta=1$, $q = 2.66 q_F$)}
    \label{fig:grid-intialization}
\end{figure}

\begin{figure}
    \centering
    \includegraphics[width=0.5\linewidth]{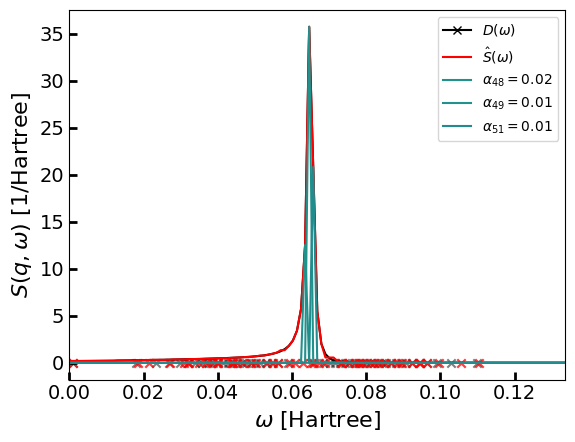}%
    \includegraphics[width=0.5\linewidth]{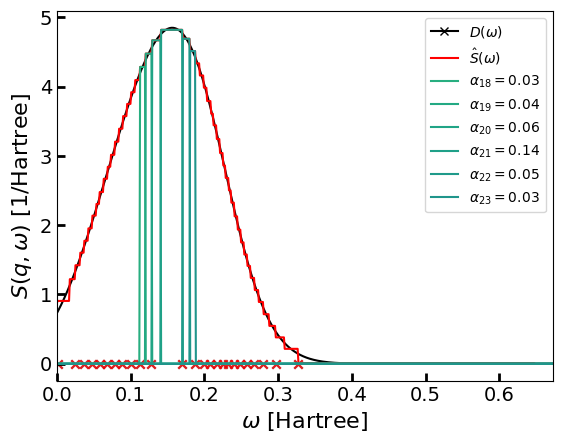}
    \caption{Plot of the final estimate of the default model obtained by the simulated annealing algorithm for uniform kernels. Left: RPA default model with $r_s=10$, $\Theta=1$, $q = 0.627 q_F$. Right: RPA default model with $r_s=10$, $\Theta=1$, $q = 2.66 q_F$.}
    \label{fig:SA_uniform_reconstructions}
\end{figure}

For completeness, we give the reconstruction of the default model obtained using uniform kernels in Fig.~\ref{fig:SA_uniform_reconstructions}. 

%\subsection{Feature: Adaptive Approach \label{app:Adaptive}}
%As presented in def. \ref{def:phys_constraints} point 3, we want there to be a sequence \((n_k)_{k \in \N}\) so that the limit value of $K_{n_k}$ is in \(\mathcal{S}\). This statement motivates the idea of choosing a sequence \(n_k\) and increasing the \(k\) until we reach a minimum.

%\begin{algorithm}
%\caption{Adaptive approach for solving the optimization %problem}\label{alg:adaptive}
%\SetKwInOut{Require}{Require}
%\SetKwInOut{Ensure}{Ensure}
%
%\Require{Precomputations:}
%minimize \(f =: f_{j_1}\) with a small conFigure \(j_1\) %\tcp*[r]{few grid points and high variance}
%minimize \(f_{j_2}\) with \(|j_2| > |j_1|\) \tcp*[r]{bigger %configuration}
%\(k \gets 2\)
%
%\While{\( \min f_{j_k} < \min f_{j_i}, \forall i < k \)}{
%    chose bigger configuration \(j_{k+1}\)\;
%    minimize \(f_{j_{k+1}}\)\;
%    \(k \gets k+1\)\;
%}
%\end{algorithm}

%\begin{algorithm}
%\caption{Adaptive approach for solving the optimization problem}\label{alg:adaptive}
%\begin{algorithmic}
%\Require Precomputations:
%\State minimize \(f =: f_{j_1}\) with a small configuration \(j_1\)
%\State\Comment{few grid points and high variance}
%\State minimize \(f_{j_2}\) with \(|j_2| > |j_1|\) \Comment{bigger configuration}
%\Ensure
%\State \(k \gets 2\)
%\While{\( \min f_{j_k} < \min f_{j_i}, \forall i < k \)}
%    \State chose bigger configuration \(j_{k+1}\)\
%    \State minimize \(f_{j_{k+1}}\)\
%    \State \(k \gets k+1\)\
%\EndWhile
%\end{algorithmic}
%\end{algorithm}

\section{Non-negative Nesterov Optimizer \label{app:nonnegative_nesterov}}
The Non-negative Nesterov is a momentum-based gradient descent method for finding the global minimum of smooth and strongly convex objective functions \(f : \mathbb{R}^n \to \mathbb{R}\)~\cite{Nesterov2012}. Given a momentum parameter \(1 > \theta_k > 0\) for \(k \in \mathbb{N}\), and a learning rate \(\eta_k > 0\), we define
\begin{align*}
    \bm{\beta}_0 &= \bm{\alpha}_0 \in \mathbb{R}^n, & & \text{(initial guess)} \\
    \bm{\beta}_k &= \max(0, (1 - \theta_k) \bm{\alpha}_k + \theta_k \bm{\alpha}_{k-1}), & \quad & \text{(non-negative momentum step)} \\
    \bm{\alpha}_{k+1} &= \bm{\beta}_k - \eta_k \nabla f(\bm{\beta}_k), & \quad & \text{(lookahead gradient step)}
\end{align*}
where the maximum is applied coordinate-wise. The lookahead mechanism reduces oscillations near the minimum thereby accelerating convergence. Choosing \(\theta_k = 2 / (k + 2)\) and \(\eta_k\) to be the Lipschitz constant \(1/\text{Lip}(f)\), Nesterov's method achieves an optimal quadratic convergence \(\mathcal{O}(1/k^2)\) for first-order methods, remaining valid also in the non-negative case.
\(\bs{E}\) denotes the matrix of kernel functions, where each row represents a kernel evaluated at the grid points \(\omega_j\). Furthermore, the Wasserstein distance can be formulated in the discrete setting using the matrix \(\bm{W} \in \mathbb{R}^{n \times n}\), defined by \(\bm{W}_{i,j} \coloneqq 1/i\) for \(j \leq i\) and zero otherwise.
% If we formulate the Wasserstein distance in a discrete sense, we introduce the following matrix representation:
% \begin{equation}\label{eq:Wasserstein_metric}
%     \bs{W} = \begin{pmatrix}
%         1 & 0 & 0 & \cdots & 0 \\
%         \frac{1}{2} & \frac{1}{2} & 0 & \cdots & 0 \\
%         \frac{1}{3} & \frac{1}{3} & \frac{1}{3} & \cdots & 0 \\
%         \vdots & \vdots & \vdots & \ddots & \vdots \\
%         \frac{1}{n} & \frac{1}{n} & \frac{1}{n} & \cdots & \frac{1}{n}
%     \end{pmatrix}
% \end{equation}
To further accelerate convergence, we choose the minimizer of the least-squares problem 
\[
    \bm{\alpha}_0 \coloneqq \arg \min_{\bm{\alpha}} \|\bs{S}_\ba - \bm{D}\|_2^2, \quad \bs{S}_\ba = \sum_j \bs{E}_{k,j} \bs{\alpha}_j,
\]
as our initial guess.

% Lastly, we set the inital coefficients $\ba_{0}$ to the coefficients that best approximate the default model. To do this, we minimize the least square problem $\|S_\ba -\bs{D}\|_2^2$ with respect to $\ba$ where $\bs{S}_\ba = \sum_j \bs{E}_{k,j} \bs{\alpha}_j$ and $\bs{E}$ is the evaluation matrix 
% \begin{equation}
%     \bs{E}_{k,j} \coloneqq K_j(\omega_k)
% \end{equation} 

% Each regularization term will produce a different $\nabla_\ba r(\cdot)$. The Lipschitz constant $L$, momentum coefficient $\gamma$, and gradient $\nabla f(\ba)$ are given for each regularizer in Table \ref{tab:NNN_implmentation}.

% derivation of GLS gradient
% \frac{1}{2} (bs{F}-\bs{R} \ba)^T   C^{-1} (bs{F}-\bs{R} \ba )
% \frac{1}{2} (bs{F}-\bs{R} \ba)^T   L L^T (bs{F}-\bs{R} \ba )
% \frac{1}{2} (L^T (bs{F}-\bs{R} \ba)^T (L^T (bs{F}-\bs{R} \ba ))
% \frac{1}{2} (bs{F*}-\bs{R*} \ba)^T (bs{F*}-\bs{R*} \ba )
% R*^T (bs{F*}-\bs{R*} \ba )

%
\begin{table}
    \caption{\label{tab:NNN_implmentation} Tabulation of non-negative Nesterov optimizer parameters.}
    \centering
    \begin{tabular}{lccc}
    \toprule
      & $r(\ba)$ & $\nabla_\ba r(\ba)$ & $\mathrm{Lip}(f)$ \\ 
    \midrule\\[-1ex]
    %     \ref{item:l1reg} &\(L^1\)-Reg.   &   \( \|\ba\|_1 \) & 1     & $\displaystyle \|\bs{R}^\top \bs{R}\|_2$ \\[3ex]
    %     \ref{item:l2reg}&\(L^2\)-Reg.   &     \(  \frac{1}{2} \| \ba\|_2^2 \) & \(\ba\)    & $\|\bs{R}^\top \bs{R}\|_2 + \lambda$ \\[3ex]
    %     \ref{item:tvreg}&TV             &    \(  \frac{1}{2} \left\| \bs{T} \ba  \right\|_{2}^2 \) & \( \bs{T}^\top (\bs{T} \ba) \)    & $\|\bs{R}^\top \bs{R}\|_2 + \lambda \|\bs{T}^\top \bs{T}\| $ \\[3ex]
    %     \ref{item:varreg}&Var            & $(\bs{M})^\top \operatorname{diag}(\bs{E}\ba)(\bs{M})  $ &  $-2 \bs{E}^\top \bs{\omega}\bs{M} \operatorname{diag}(\bs{E}\ba) + \bs{E}^\top \operatorname{diag}\bs{M}^2. $ & $\|\bs{R}^\top \bs{R}\|_2 + \lambda \|\bs{E}\bs{\omega} \|_2 \|\bs{E}^\top\|_2 \|\bs{\omega}\|_2 $ \\[3ex]
     \(L^2\)-Fit    &   \( \frac{1}{2}\|\bs{E}\ba-\bs{D}\|_{2}^2\) & \(\ \bs{E}^\top(\bs{E}\ba-\bs{D})\)    & $\|\bs{R}^\top \bs{R}\|_2 + \lambda \|\bs{E}^\top \bs{E}\|$ \\[3ex]
     CE             &    \(\displaystyle - (\bs{E}\ba)^\top \ln \frac{\bs{E}\ba}{\bs{D}} \) & \( \displaystyle -\bs{E}^\top \left(\log \frac{\bs{E} \ba}{\bs{D}} + \bs{1}\right) \)   & $\|\bs{R}^\top \bs{R}\|_2 + \lambda \|\bs{E}\|_2^2$ \\[3ex]
     WD   &    \( \frac{1}{2} (\bs{E}\ba - \bs{D})^\top \bs{W} (\bs{E}\ba - \bs{D}) \) & \( \bs{E}^\top \bs{W} (\bs{E}\ba - \bs{D})\)    & $\| \bs{R}^\top \bs{R} + \lambda \bs{E}^\top \bs{W} \bs{E} \|_2 $ \\[1ex]
     \bottomrule
\end{tabular}
\end{table}

\section{Automatic regularization weight \texorpdfstring{$\lambda$}~ selection\label{app:lambda_selection}}
The regularization weight $\lambda$ in our optimization problem \eqref{eq:min_problem} is not specified by PyLIT. Since this value affects the solution it introduces uncertainty and must be considered carefully. Varying $\lambda$ gives rises to a typical shape of the fidelity term $\frac{1}{2}\|\bF - \bs{R} \ba\|_2^2$ in \eqref{eq:min_problem}, which resembles a logistic growth curve. As $\lambda \rightarrow 0$, the fidelity dominates the optimization problem and achieves its minimum as the solution converges the least squares solution; this is called the ``noise-fitting'' region. As $\lambda \rightarrow \infty$, the fidelity term is unimportant in the optimization and and achieves its maximum as the solution converges to the default model; this is called the ``default model'' region. In between these two limits there is an ``information fitting'' region where the fidelity value typically changes logistically between these limits. Typically, authors prefer to select regularization weight $\lambda$ in the information fitting region.

There exist many different approaches to selecting the regularization weight. First, Gull's Bayesian posterior \cite{gull1989MEMBayesianWeighting}, commonly used in the maximum entropy method \cite{bryan1990algorithm, JARRELL1996133, Asakawa2001MEM}, assumes that there is a Gaussian distribution of acceptable $\lambda$ values centered about the $\lambda$ value which minimizes the cost function and then averages over this collection of solutions. Second, Kaufmann and Held's $\chi^2$-kink method, which was developed in response to Gull's Bayesian procedure. This method posits that the best regularization weight is at the very start of the information region \cite{KAUFMANN2023ana_cont} and introduces an \textit{ad hoc} parameter to select in that region. Kaufmann and Held major reasoning here being that Gull's selected $\lambda$ values were empirically observed to be in the noise fitting region. Next, we detail both approaches and our implementations.

Gull's Bayesian posterior is given by
\begin{align}
    \langle S(q,\omega) \rangle = \int \mathrm{d}\lambda  P[\lambda \,| F, D ] S_\lambda(q,\omega).
\end{align}
The Bayesian posterior weighting function $P[\lambda| F, D ]$, given as
\begin{align}\label{eq:solutionweight}
        P[\lambda \mid F, D] = \exp\left( - \frac{1}{2} \chi^2(S_\lambda) + \lambda \,  r(S_\lambda, D) + \frac{1}{2} \sum_i \log \left( \frac{\lambda}{\lambda + \lambda_i} \right) \right).
\end{align}
Here $\lambda_i$ is the typical $i^\mathrm{th}$ eigenvalue of the real symmetric matrix $\Lambda_{j,j'}(\alpha) = \sqrt{x_j(\alpha)} \frac{\partial^2 L}{\partial x_j(\alpha) \partial x_{j'}(\alpha)} \sqrt{x_{j'}(\alpha)}$. However, our weighting function uses a generalized $\lambda \,  r(S_\lambda, D)$, instead of the typical entropic term that is used in the maximum entropy method \cite{gull1989MEMBayesianWeighting, bryan1990algorithm, JARRELL1996133}. This substitution is fine as long as the cost function $- \frac{1}{2} \chi^2(S_\lambda) + \lambda \,  r(S_\lambda, D)$ has a unique minimum about which the Gaussian approximation can be made. Thus, in this work we use a generalized version of Gull's posterior. 

In practice, we conduct a logarithmic grid search over the regularization weight $\lambda$. Then, following Asakwa et al. \cite{Asakawa2001MEM}, we take a weighted combination over a subdomain $[\lambda_\textrm{min}, \lambda_\textrm{max}]$ where the posterior distribution $P[\lambda \mid F, D]$ and the maximum of that distribution $P[\lambda^* \mid F, D]$ are related by the bound $P[\lambda  \mid F, D] \leq 0.1 \times P[\lambda^* \mid F, D]$. 
The final estimate $\langle x \rangle$ of the DSF is determined from the collection of solutions by
\begin{align}
     \langle S_\lambda(q,\omega) \rangle = \frac{\sum_{\lambda, } S_\lambda(q,\omega) \, P[\lambda \mid F, D]}{ \sum_\lambda P[\lambda \mid F, D]},
\end{align}
where the indices $\lambda$ are the $\lambda$ values within $[\lambda_\textrm{min}, \lambda_\textrm{max}]$ and $S_\lambda$ is the corresponding solution. Our uncertainty in the solution is then given by the variance $V[S_\lambda] = \langle (S_\lambda)^2 \rangle - \langle S_\lambda \rangle^2$.

Kaufmann and Held's logistic growth curve is defined
\begin{align}
    f(\lambda \mid \vec{p}) = p_0 + \frac{p_1}{ 1 + \exp(-p_3(\lambda - p_2))}
\end{align}
and fit to the logarithm of $\lambda$ and the fidelity as 
\begin{align}
    \log(\chi^2(\lambda)) = f(\log(\lambda))\ .
\end{align}
Then we select a $\lambda$ near where the second derivative has its max value using
\begin{align}
    \lambda_\text{best} = 10^{p_2 - p_\chi /p_3}\ .
\end{align}
Here the $p_\chi$ is the \textit{ad hoc} parameter that the authors included because they wanted to select $\lambda_\text{best}$ values at smaller $\lambda$ than where the second derivative is largest; they recommend $p_\chi \in [2,2.5]$. 

In practice, we conduct a logarithmic grid search over the regularization weight $\lambda$ and fit a logarithmic growth curve to the fidelity term. Then we estimate the solution for $p_\chi$ parameter from $2-2.5$. The average and variance from this collection of solutions constitutes the estimate and its error. 

\begin{figure}
    \centering
    \includegraphics[width=0.5\linewidth]{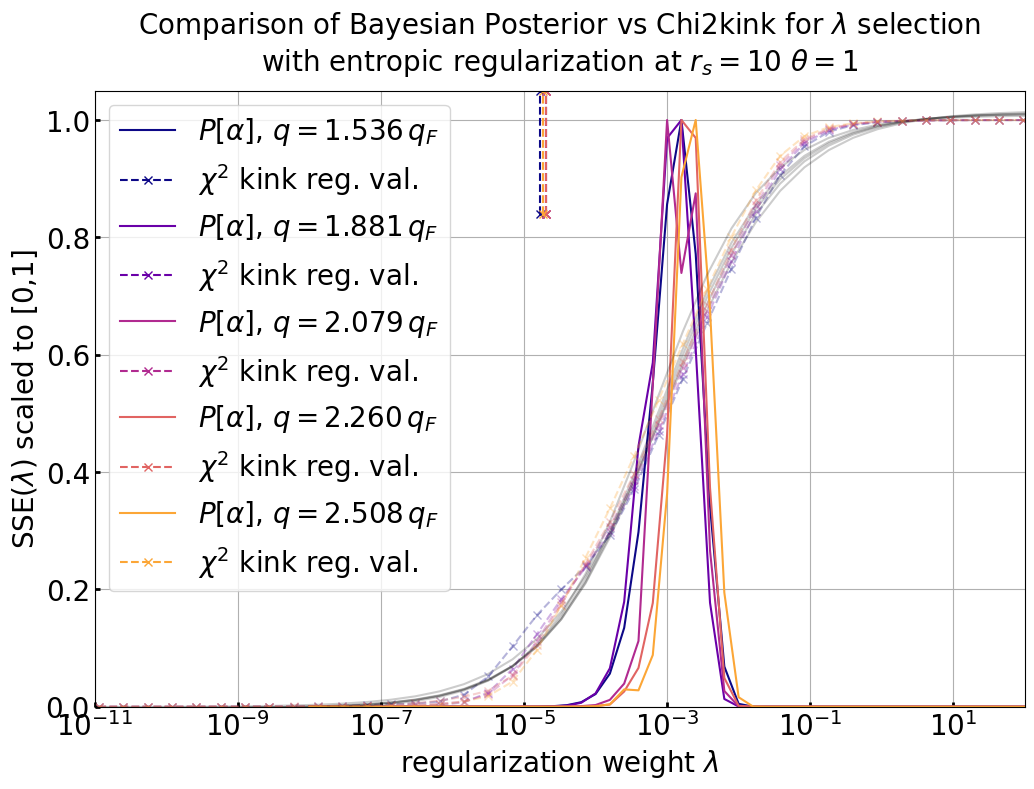}
        \includegraphics[width=0.5\linewidth]{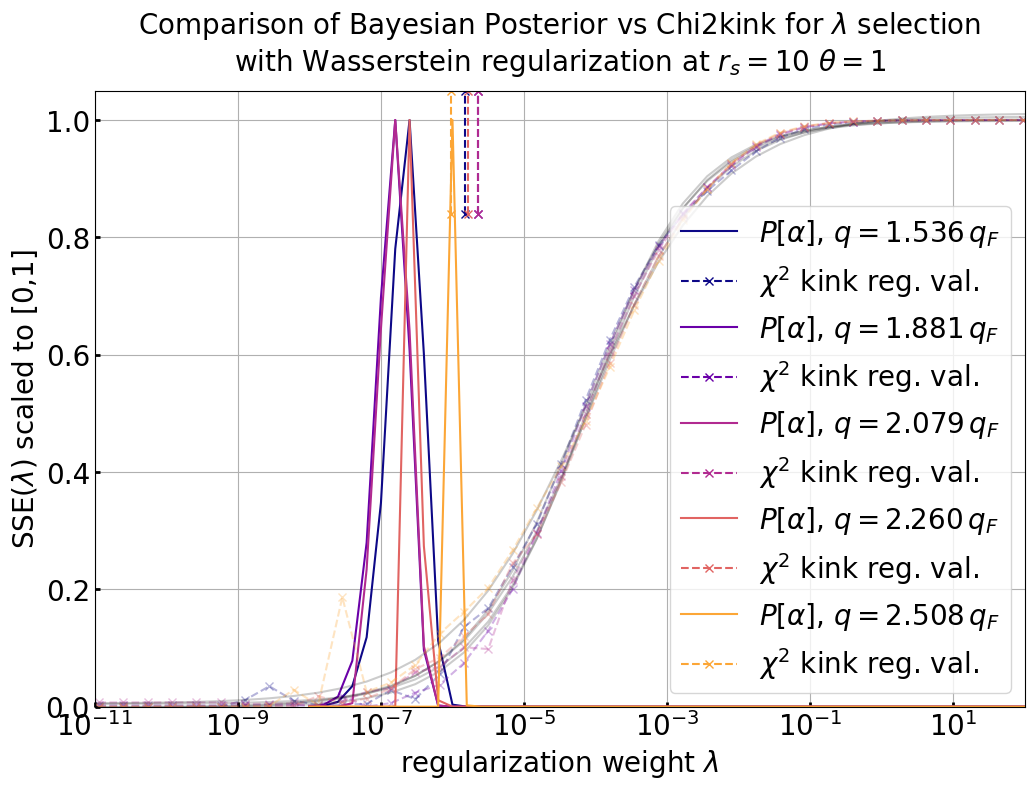}
    \includegraphics[width=0.5\linewidth]{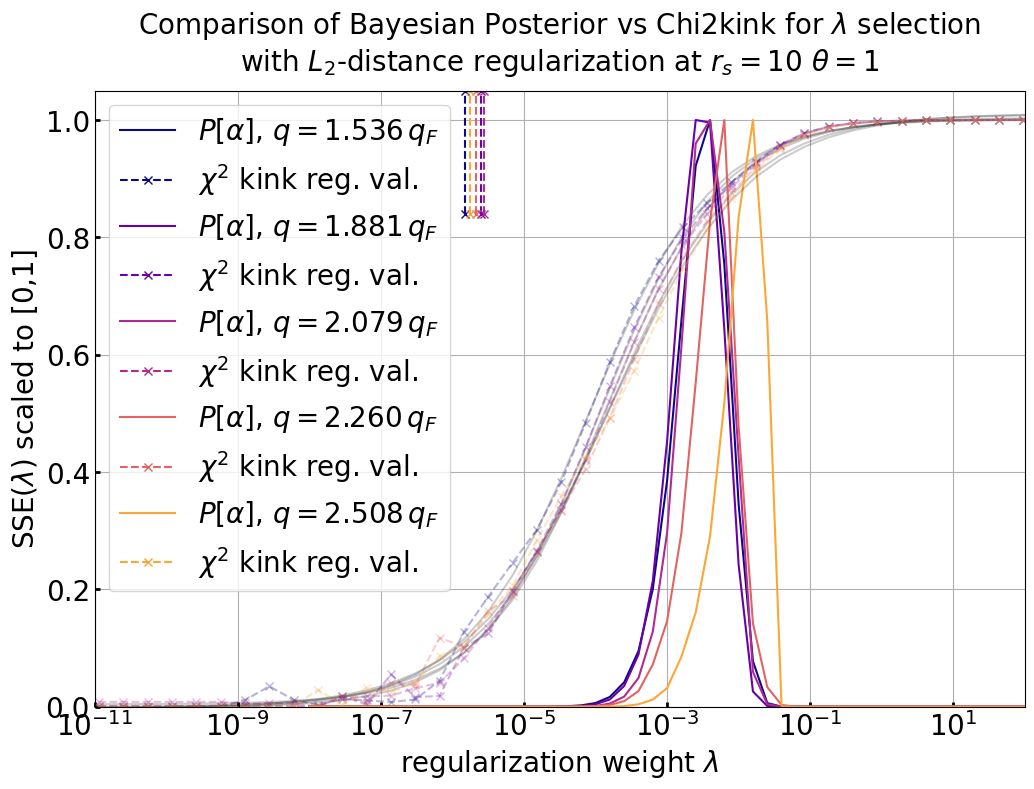}
    \caption{Plots of the Gull Bayesian posterior where the max value is normalized to 1 compared to the fidelity term as a function of $\lambda$ scaled from $[0,1]$ and the $\lambda$ value selected by the $\chi^2$-kink algorithm. We plot these curves for optimizations conducted at 5 $q$ values listed in the legend. Each curve is obtained from synthetic data at $r_s=10$, $\theta=1.0$, and noise level $\sigma = 0.01$ using 50 Gaussian kernels; different plots use different regularization terms. In these plots, we see the typical noise-fitting region for $\lambda \rightarrow 0$, the default model region for $\lambda \rightarrow \infty$, and the information region for intermediate $\lambda$. We see that both Gull's Bayesian weighting and the $\chi^2$-kink algorithm select values within the information region, but Gull's procedure is more responsive to the choice of the regularization.}
    \label{fig:chi2kinkBayesianComparison}
\end{figure}

We compare of the methods for 50 Gaussian kernel on synthetic data at $r_s=10$ $\theta=1.0$, noise level $\sigma = 0.01$. We plot Gull's Bayesian posterior and Kaufman and Held's  $chi^2$-kink selection and the logistic growth of the fidelity term in Fig.~\ref{fig:chi2kinkBayesianComparison}.

\section{Normalization and time scaling \label{app:tau_scaling}}
For avoiding numerical overflow, we re-scale the model, $\widehat u \circ \varphi$ with respect to the scaling, formally given by $\varphi : \R \to \R$, with 
\begin{equation}
    \varphi(\tau) = \frac{\tau}{\widehat\tau}\,,
\end{equation}
where \(\widehat\tau =\max(\tau)\) ist the $\tau$-domain boundary. 

The re-scaling of the input data follows the same logic. 
%\todo{This only refers to the scaling of the model. There should be a small text here explaining what happens with the input}
% maybe lets refer to dividing by max F and integral of S as Normalization and only applying phi as scaling
Note that the re-scaled Laplace transformation is given by
\begin{equation}
     \widehat u \circ \varphi 
    = \int_\infty^{-\infty}  \e^{- \tau \omega}  \widehat\tau   u(\widehat\tau\omega)\,\mathrm{d}{\omega}.
\end{equation}
This results the inverse of the re-scaled Laplace transformation as
\begin{equation}
    \mathcal{L}^{-1}\{\widehat u \circ \varphi\} =  \widehat{\tau} u(\widehat\tau\omega)\,.
\end{equation}

When re-scaling the input, we assume the Laplace transform at $\tau=0$ to be given as 
\begin{equation}
    \mathcal{L}\{S_\ba(\omega)\}[0]=\int_{-\infty}^\infty S_\ba(\omega) \dom = 1\,.
\end{equation}
Consequently,  the input \(\bs{F}\)  itself must be re-scaled accordingly. Thus, \(\bs{F}\) becomes \(\bs{F}/\bs{F}_0 = \bs{F}/\max(\bs{F})\), which is reversed for the output.

Albeit explicitly outlined above, we consider re-scaling as a black box, being computed in the background, with unchanged output-scale. 

\bibliography{bibliography}

\end{document}